\documentclass[iop,apj,times]{emulateapj}

\usepackage{natbib}
\usepackage{color}
\usepackage{ulem}

\newcommand{\rev}{\textcolor{black}}
\newcommand{\proof}{\textcolor{black}}

\begin{document}

\title{OGLE-2012-BLG-0563Lb: a Saturn-mass Planet around an M Dwarf with the Mass Constrained by {\it Subaru} AO imaging}

\author{
% lead authors ------------------------------
A.~Fukui$^{M1, M}$, A.~Gould$^{U1, U}$, T.~Sumi$^{M2, M}$,
D.~P.~Bennett$^{M3, M}$, I.~A.~Bond$^{M4, M}$, C.~Han$^{U2, U}$,
D.~Suzuki$^{M3, M}$, J.-P.~Beaulieu$^{P1, P}$, V.~Batista$^{P1, U, P}$,
A.~Udalski$^{O1, O}$, R.~A.~Street$^{R1, R}$, Y.~Tsapras$^{R1,R2,R3, R}$,
M. Hundertmark$^{R6,R7, R}$,
\\
and\\
% MOA ------------------------------
F.~Abe$^{M5}$, 
A.~Bhattacharya$^{M3}$, 
%C.~S.~Botzler$^{M6}$, 
M.~Freeman$^{M6}$, 
%D.~Fukunaga$^{M5}$, 
Y.~Itow$^{M5}$, 
C.~H.~Ling$^{M4}$, N.~Koshimoto$^{M2}$, K.~Masuda$^{M5}$,
Y.~Matsubara$^{M5}$, Y.~Muraki$^{M5}$, 
K.~Ohnishi$^{M8}$, L.~C.~Philpott$^{M9}$, N.~Rattenbury$^{M6}$,
T.~Saito$^{M10}$, D.~J.~Sullivan$^{M7}$, P.~J.~Tristram$^{M11}$,
A.~Yonehara$^{M12}$\\
%P.~C.~M.~Yock$^{M6}$\\
(The MOA Collaboration),\\
% uFUN ------------------------------
%.~A.~Almeida$^{U8}$,
%G.~Christie$^{U3}$,
%J.-Y.~Choi$^{U2}$,
%D.~L.~DePoy$^{U4}$,
%B.~S.~Gaudi$^{U1}$,
%C.~Henderson$^{U1}$,
%K.-H.~Hwang$^{U2}$,
%F.~Jablonski$^{U8}$,
%Y.~K.~Jung$^{U2}$,
%C.-U.~Lee$^{U5}$,
%J.~McCormick$^{U6}$,
%T.~Natusch$^{U3,U7}$,
%H.~Ngan$^{U3}$,
%H.~Park$^{U2}$,
%R.~W.~Pogge$^{U1}$,
%I-G.~Shin$^{U2}$,
%J.~C.~Yee$^{U1,U9}$\\
J.-Y.~Choi$^{U2}$, G.W.~Christie$^{U3}$, D.L.~DePoy$^{U4}$, 
Subo~Dong$^{U5}$, J.~Drummond$^{U6}$, B.S.~Gaudi$^{U1}$, 
K.-H.~Hwang$^{U2}$, A.~Kavka$^{U1}$, C.-U.~Lee$^{U7}$,
J.~McCormick$^{U8}$, T.~Natusch$^{U3, U9}$, H.~Ngan$^{U3}$,
H.~Park$^{U2}$, R.W.~Pogge$^{U1}$, I-G.~Shin$^{U2}$,
T.-G.~Tan$^{U10}$, J.C.~Yee$^{U1,U11,U12}$\\
(The $\mu$FUN Collaboration)\\
% OGLE -----------------------------
M.~K.~Szyma{\'n}ski$^{O1}$, G.~Pietrzy{\'n}ski$^{O1}$,
I.~Soszy{\'n}ski$^{O1}$, R.~Poleski$^{O1,U1}$, S.~Koz{\l}owski$^{O1}$,
P.~Pietrukowicz$^{O1}$, K.~Ulaczyk$^{O1}$, {\L}.~Wyrzykowski$^{O1}$\\
(The OGLE Collaboration),\\
% RoboNet --------------------------
D. M. Bramich$^{R4}$,
P. Browne$^{R5}$,
M. Dominik$^{R5, \dagger}$,
K. Horne$^{R5}$,
S. Ipatov$^{R8,R9}$,
N. Kains$^{R10,R11}$,
C. Snodgrass$^{R12,R13}$,
I. A. Steele$^{R14}$\\
(The RoboNet Collaboration)\\
}
% -----------------------------------------------------------------------------------------------------------
\affil{$^{M1}$ Okayama Astrophysical Observatory, National Astronomical Observatory of Japan, Asakuchi, 719-0232 Okayama, Japan} 
\affil{$^{M2}$ Dept. of Earth and Space Science, Graduate School of Science, Osaka University, 1-1 Machikaneyama-cho, Toyonaka, 560-0043 Osaka, Japan} 
\affil{$^{M3}$ Dept. of Physics, University of Notre Dame, Notre Dame, IN 46556, USA} 
\affil{$^{M4}$ Institute of Information and Mathematical Sciences, Massey University, Private Bag 102-904, North Shore Mail Centre, Auckland, New Zealand} 
\affil{$^{M5}$ Solar-Terrestrial Environment Laboratory, Nagoya University, 464-8601 Nagoya, Japan} 
\affil{$^{M6}$ Dept. of Physics, University of Auckland, Private Bag 92019, Auckland, New Zealand} 
\affil{$^{M7}$ School of Chemical and Physical Sciences, Victoria University, Wellington, New Zealand} 
%\affil{$^{M8}$ Dept. of Physics, Konan University, Nishiokamoto 8-9-1, 658-8501 Kobe, Japan} 
\affil{$^{M8}$ Nagano National College of Technology, 381-8550 Nagano, Japan} 
\affil{$^{M9}$ Department of Earth, Ocean and Atmospheric Sciences, University of British Columbia, Vancouver, British Columbia, V6T 1Z4, Canada}
\affil{$^{M10}$ Tokyo Metropolitan College of Industrial Technology, 116-8523 Tokyo, Japan} 
\affil{$^{M11}$ Mt. John University Observatory, P.O. Box 56, Lake Tekapo 8770, New Zealand}
\affil{$^{M12}$ Department of Physics, Faculty of Science, Kyoto Sangyo University, Motoyama, Kamigamo, Kita-Ku, Kyoto, Kyoto 603-8555, Japan}
% ===========================================================
%\affil{$^{U1}$ Department of Astronomy, Ohio State University, 140 West 18th Avenue, Columbus, OH 43210, USA}
%\affil{$^{UH2}$ Department of Physics, Institute for Astrophysics, Chungbuk National University, 371-763 Cheongju, Korea}
%\affil{$^{U3}$ Auckland Observatory, 670 Manukau Rd, Royal Oak 1023, Auckland, New Zealand}
%\affil{$^{U4}$ Dept. of Physics and Astronomy, Texas A\&M University College Station, TX 77843-4242, USA}
%\affil{$^{UH5}$ Korea Astronomy and Space Science Institute, 305-348 Daejeon, Korea}
%\affil{$^{U6}$ Farm Cove Observatory, Centre for Backyard Astrophysics, Pakuranga, Auckland, New Zealand}
%\affil{$^{U7}$ Institute for Radiophysics and Space Research, AUT University, Auckland, New Zealand}
%\affil{$^{U8}$ Divis\~{a}o de Astrofisica, Instituto Nacional de Pesquisas Espeaciais, Avenida dos Astronauntas, 1758 Sao Jos\'e dos Campos, 12227-010 SP, Brazil}
%\affil{$^{U9}$ Harvard-Smithsonian Center for Astrophysics, 60 Garden St., Cambridge, MA 02138, USA}
%
\affil{$^{U1}$ Department of Astronomy, Ohio State University, 140
W.\ 18th Ave., Columbus, OH 43210, USA;
gould@astronomy.ohio-state.edu}
%\affil{$^{U2}$ Institut d'Astrophysique de Paris, Universit'e Pierre et Marie Curie, CNRS UMR7095, 98bis Boulevard Arago, 75014 Paris, France}
\affil{$^{U2}$ Department of Physics, Institute for Astrophysics, Chungbuk National University, 371-763 Cheongju, Korea}
\affil{$^{U3}$ Auckland Observatory, Auckland, New Zealand;
gwchristie@christie.org.nz}
\affil{$^{U4}$ Dept.\ of Physics, Texas A\&M University, College Station, TX, USA;
depoy@physics.tamu.edu}
\affil{$^{U5}$ Kavli Institute for Astronomy and Astrophysics,
Peking University, Yi He Yuan Road 5, Hai Dian District, Beijing
100871, China}
\affil{$^{U6}$ Possum Observatory, Patutahi,
New Zealand}
\affil{$^{U7}$ Korea Astronomy and Space Science Institute, 305-348 Daejeon, Korea}
\affil{$^{U8}$ Farm Cove Observatory, Centre for Backyard Astrophysics,
Pakuranga, Auckland, New Zealand; farmcoveobs@xtra.co.nz}
\affil{$^{U9}$ AUT University, Auckland, New Zealand; tim.natusch@aut.ac.nz}
\affil{$^{U10}$ Perth Exoplanet Survey Telescope, Perth, Australia}
\affil{$^{U11}$ Harvard-Smithsonian Center for Astrophysics, 60 Garden St., Cambridge, MA 02138, USA}
\affil{$^{U12}$ Sagan Fellow}
% ===========================================================  
\affil{$^{O1}$ Warsaw University Observatory, Al. Ujazdowskie 4, 00-478 Warszawa, Poland}
%\affil{$^{O2}$ Universidad de Concepci\'on, Departamento de Astronomia, Casilla 160-C, Concepci\'on, Chile}
%\affil{$^{O3}$ Institute of Astronomy, University of Cambridge, Madingley Road, Cambridge CB3 0HA, UK}
% ===========================================================
\affil{$^{P1}$ Institut d'Astrophysique de Paris, Universit'e Pierre et Marie Curie, CNRS UMR7095, 98bis Boulevard Arago, 75014 Paris, France}
\affil{$^{R1}$Las Cumbres Observatory Global Telescope Network, 6740 Cortona Drive, suite 102, Goleta, CA 93117, USA}
\affil{$^{R2}$Astronomisches Rechen-Institut, Zentrum f{\"u}r Astronomie der Universit{\"a}t Heidelberg (ZAH), 69120 Heidelberg, Germany} 
\affil{$^{R3}$School of Physics and Astronomy, Queen Mary University of London, Mile End Road, London E1 4NS, UK}
\affil{$^{R4}$Qatar Environment and Energy Research Institute, Qatar Foundation, P.O. Box 5825, Doha, Qatar}
\affil{$^{R5}$SUPA, School of Physics \& Astronomy, University of St Andrews, North Haugh, St Andrews KY16 9SS, UK}
\affil{$^{R6}$Niels Bohr Institute, University of Copenhagen, Juliane Maries Vej 30, 2100, K{\o}benhavn {\O}, Denmark}
\affil{$^{R7}$Centre for Star and Planet Formation, Natural History Museum, University of Copenhagen, {\O}stervoldgade 5-7, 1350, K{\o}benhavn K, Denmark}
\affil{$^{R8}$Vernadsky Institute of Geochemistry and Analytical Chemistry of Russian Academy of Sciences, Kosygina 19, 119991, Moscow, Russia}
\affil{$^{R9}$Space Research Institute of Russian Academy of Sciences, Profsoyuznaya st. 84/32, Moscow, Russia}
\affil{$^{R10}$European Southern Observatory, Karl-Schwarzschild-Str. 2, 85748 Garching bei M\"unchen, Germany}
\affil{$^{R11}$Space Telescope Institute, 3700 San Martin Drive, Baltimore, MD 21218, USA}
\affil{$^{R12}$Planetary and Space Sciences, Department of Physical Sciences, The Open University, Milton Keynes, MK7 6AA, UK}
\affil{$^{R13}$Max Planck Institute for Solar System Research, Justus-von-Liebig-Weg 3, 37077 G\"{o}ttingen, Germany}
\affil{$^{R14}$Astrophysics Research Institute, Liverpool John Moores University, Liverpool L3\ 5RF, UK}
\affil{$\dagger$ Royal Society University Research Fellow}
\footnotetext [M] {Microlensing Observations in Astrophysics (MOA) Collaboration.}
\footnotetext [U] {Microlensing Follow-up Network ($\mu$FUN) Collaboration.}
\footnotetext [O] {Optical Gravitational Lensing Experiment (OGLE) Collaboration.}
\footnotetext [P] {Probing Lensing Anomalies NETwork (PLANET) Collaboration.}
\footnotetext [R] {RoboNet Collaboration.}
%\affil{$^{S}$ The MiNDSTEp Consortium}

\begin{abstract}
We report the discovery of a microlensing exoplanet OGLE-2012-BLG-0563Lb with the planet-star mass ratio of $\sim1\times10^{-3}$.  Intensive photometric observations of a high-magnification microlensing event allow us to detect a clear signal of the planet. 
Although no parallax signal is detected in the light curve, we instead succeed at detecting the flux from the host star in high-resolution $JHK'$-band images obtained by the {\it Subaru}/AO188 and Infrared Camera and Spectrograph instruments, allowing us to constrain the absolute physical parameters of the planetary system. 
With the help of spectroscopic information about the source star obtained during the high-magnification state by Bensby et al., 
we find that the lens system is located at 1.3~$^{+0.6}_{-0.8}$ kpc from us, and consists of an M dwarf (0.34~$^{+0.12}_{-0.20}$~M$_\odot$) orbited by a Saturn-mass planet (0.39~$^{+0.14}_{-0.23}$~M$_\mathrm{Jup}$) at the projected separation of 0.74~$^{+0.26}_{-0.42}$ AU (close model) or 4.3~$^{+1.5}_{-2.5}$ AU (wide model). 
The probability of contamination in the host star's flux, which would reduce the masses by a factor of up to three, is estimated to be 17~\%. This possibility can be tested by future high-resolution imaging.
We also estimate the $(J-K_\mathrm{s})$ and $(H-K_\mathrm{s})$ colors of the host star, which are marginally consistent with a low metallicity mid-to-early M dwarf, although further observations are required for the metallicity to be conclusive.
This is the fifth sub-Jupiter-mass ($0.2<m_\mathrm{p}/M_\mathrm{Jup}<1$) microlensing planet around an M dwarf with the mass well constrained.  The relatively rich harvest of sub-Jupiters around M dwarfs is contrasted with a possible paucity of $\sim$1--2 Jupiter-mass planets around the same type of star, which can be explained by the planetary formation process in the core-accretion scheme.

\end{abstract}


\section{Introduction}

Microlensing is a unique and powerful technique to probe exoplanets with a wide range of masses just beyond the snow line, where gas-giant planets can efficiently form according to the core-accretion models \citep[e.g.,][]{1996Icar..124...62P,2002ApJ...581..666K}. The planetary-mass distribution probed by microlensing therefore provides valuable information about the planetary formation process, which is less affected by several post-formation effects such as orbital migration and mass loss due to stellar irradiation. In addition, microlensing is most sensitive to exoplanets around M dwarfs including late-type ones, which have not sufficiently been surveyed by other detection techniques due to their faintness. The core-accretion models predict that massive Jovian planets are rare around low-mass stars due to the lack of planet-forming materials \citep[e.g.,][]{2005ApJ...626.1045I}, which can thus be tested by microlensing.

Thanks to a huge effort by microlensing surveys and follow-up projects to date, 
the number of microlensing planets has reached 35,\footnote{http://exoplanet.eu} among which $\sim60$\% are hosted by M dwarfs.
These discoveries have revealed that low-mass planets are much more abundant than massive ones, which is in agreement with the core-accretion scenarios \citep{2010ApJ...720.1073G,2010ApJ...710.1641S,2012Natur.481..167C}. On the other hand, super-Jupiter-mass planets ($\gtrsim2 M_\mathrm{Jup}$) have also been discovered around M dwarfs \citep[e.g.,][]{2009ApJ...695..970D,2011A&A...529A.102B,2014ApJ...782...48T}, which challenges the same scenarios.

However, the statistics of microlensing planets are not yet high enough to draw a clear structure of the planetary-mass distribution, in terms of number and accuracy.
In particular, about half of all planetary microlensing events do not show parallax effects in the light curves, without which one cannot measure the absolute masses of the planet and host star from the light curve alone. In such cases, the physical parameters of the planetary system have often been estimated by the Bayesian technique, which uses Galactic-model priors (a stellar-mass function, stellar number density, and stellar velocity distribution) to draw posterior probability distributions of the physical parameters. This technique could be meaningful if the number of planets is statistically large enough, however, the individual values are not accurate. Furthermore, this technique relies on an assumption that the planet occurrence probability is uniform for all stars independent of stellar properties, such as stellar mass and Galactic location, and therefore the results should be treated with caution.

Another method to constrain the physical parameters of the lens system is detecting (or putting an upper limit on) the emission from the host (lens) star by high-resolution imaging. High resolution is essential to de-blend unrelated stars and extract the lens+source composite flux.
Although, for the current facilities, it is usually not possible to spatially resolve the lens star from the background source star until a decade after the microlensing event, even without resolving the two stars, the lens star's flux can be extracted by subtracting the source star's flux (obtained by a light-curve analysis) from the lens+source composite flux. The extracted lens flux provides a mass-distance relation of the host star, allowing us to solve for the mass and distance by combining with another mass-distance relation provided by the angular Einstein radius $\theta_\mathrm{E}$, which can be derived in most planetary microlensing events.

High-resolution imaging has now become an important tool to constrain the physical parameters of the lens systems. 
So far, 11 planetary events have been imaged at high resolution using {\it the Hubble Space Telescope} \citep[e.g.,][]{2006ApJ...647L.171B} or ground-based adaptive-optics (AO) instruments \citep[Very Large Telescope; VLT/NACO or Keck/NIRC2; e.g.,][]{2010ApJ...711..731J,2010ApJ...713..837B}. Among them, five events did not show parallax effects in their light curves, and therefore  the high-resolution images played a complementary role to constrain the lens-system parameters.
Regarding M-dwarf host star events, the following six have been imaged to date with high resolution: OGLE-2005-BLG-071 \citep{2005ApJ...628L.109U,2009ApJ...695..970D}, OGLE-2006-BLG-109 \citep{2008Sci...319..927G,2010ApJ...713..837B}, MOA-2007-BLG-192 \citep{2008ApJ...684..663B,2012A&A...540A..78K}, and MOA-2009-BLG-387 \citep{2011A&A...529A.102B}, OGLE-2013-BLG-0341 \citep[][Batista et al. in preparation]{2014Sci...345...46G}, and MOA-2011-BLG-262 \citep{2014ApJ...785..155B}. Among them, the first five events also showed parallax signals in the light curves, and thus the high-resolution images have been used to support or reinforce the results from the light-curve analyses. As for the last event, MOA-2011-BLG-262, the lens flux was not clearly detected. This event has actually two degenerate solutions: a planet-M-dwarf system in the Galactic bulge and a nearby free-floating planet orbited by a moon.

In this paper, we report the discovery of a new planetary microlensing event with the planet-star mass ratio of $\sim10^{-3}$ without parallax signal, for which we constrain the physical parameters of the lens system by combining light-curve analysis with {\it Subaru}/AO imaging. This is the first M-dwarf-host planetary event without parallax for which the lens flux is clearly detected, and therefore the AO imaging plays a crucial role for deriving the physical parameters.

The rest of this paper is organized as follows. The observations and light-curve analysis are described in Sections \ref{sec:obs} and \ref{sec:lc-anal}, respectively. The properties of the source star are investigated in Section \ref{sec:CMD}. Sections \ref{sec:extracting_excess_flux} and \ref{sec:physical_parameters} describe the extraction of excess flux on the source star and constraint on the physical parameters of the lens system, respectively. We discuss the results in Section \ref{sec:discussion} and summarize in Section 8.%\ref{sec:summary}.

\section{Observations}
\label{sec:obs}

\subsection{The Microlensing Event}
The microlensing event OGLE-2012-BLG-0563 occurred on a star located at the equatorial coordinate of ($\alpha$, $\delta$)$_\mathrm{J2000}$ = (18$^\mathrm{h}$05$^\mathrm{m}$57$^\mathrm{s}$.72, -27$^\circ$42$'$43\farcs2) and the Galactic coordinate of ($l$, $b$) = (3\fdg31, -3\fdg25). This event  was first discovered by the OGLE collaboration on 2012 May 1 UT ($\mathrm{JD}' \equiv \mathrm{JD}-2450000=6049$), during its regular photometric monitoring toward the Galactic bulge by using the 1.3m Warsaw telescope equipped with the wide field (1.4deg$^2$) camera OGLE-IV at Las Campanas Observatory in Chile \citep{2015AcA....65....1U}. The event was discovered in the field BLG519, which was regularly monitored twice a night with an $I$-band filter.
The same event was independently discovered as MOA-2012-BLG-288 by the MOA collaboration on 2012 May 18 UT ($\mathrm{JD}'=6066$), by using the 1.8m MOA-II telescope and the wide field (2.2-deg$^2$) camera moa-cam3 \citep{2008ExA....22...51S} at Mt. John University Observatory (MJUO) in New Zealand. MOA monitored the event  (in field gb14) with a typical cadence of 20 minutes using a custom $R+I$ filter. After these discoveries, both teams have made the light curves public and updated them frequently.

On May 20 ($\mathrm{JD}'=6067.9$), the $\mu$FUN collaboration circulated an alert that the event was peaking at a high magnification,
which means that the source star was approaching very close to the lens star on the sky plane.
In such a case, there is a high probability that the light curve will show anomalous features around the peak if the lens star hosts a planet, because 
 one of the caustics (central caustic) produced by a planetary system is always created near the host star on the sky plane \citep{1998ApJ...500...37G,2000ApJ...533..378R,2001ApJ...546..975H}.

\begin{figure}
\begin{center}
\includegraphics[width=9cm]{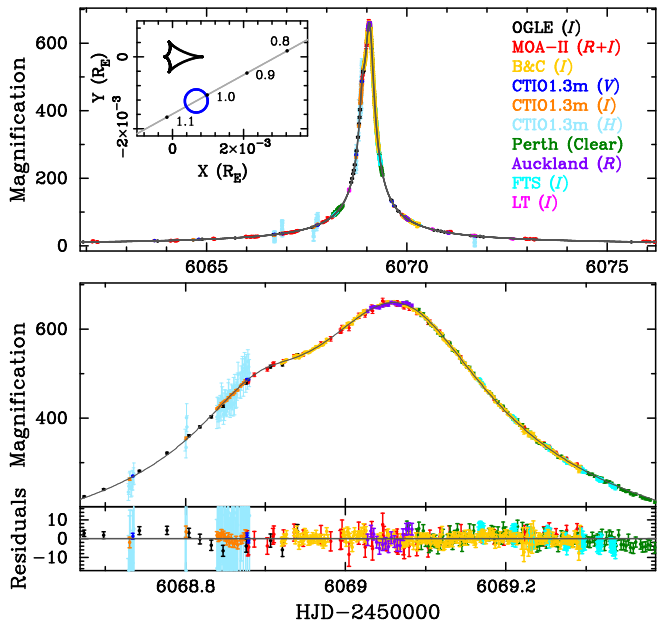}
\caption{
(Top) light curve of the event OGLE-2012-BLG-0563. The correlation between color and instrument is shown on the right. The solid black line indicates the best-fit planetary microlensing model. 
The inset shows the geometry of the event projected onto the lens plane. The origin is the centroid of the planetary system, and the $X$ axis is defined as the host star-planet axis. The black and gray lines indicate the central caustic for the best-fit close model (which is almost identical to that for the wide model) and the source trajectory, respectively. The blue circle represents the source star at $t_0$, with the size showing the source size. The black dots are tick marks for time, with number indicating HJD-2456068.     (Middle) a zoom around the peak. (Bottom) residuals of the zoomed light curve from the best-fit model.
}
\label{fig:lc_all}
\end{center}
\end{figure}

After the circulation of the high-magnification alert, the $\mu$FUN and RoboNet collaborations started to follow up the event with high cadence photometry. $\mu$FUN used the following telescopes (filters): the 1.3m CTIO telescope in Chile ($V$, $I$, and $H$), the 0.40m Auckland telescope in New Zealand ($R$), the 0.36m Possum telescope in New Zealand ($R$), the 0.36m Farm Cove telescope in New Zealand (clear filter), and the 0.30m PEST in Australia (clear filter).  The 1.3m CTIO's $H$-band data were simultaneously obtained with the V- or $I$-band data with the same instrument. RoboNet used three 2.0m telescopes, including the Liverpool Telescope (LT) in Canary Islands, Spain ($i'$); the Faulkes Telescope North (FTN) in Hawaii, USA ($i'$); and the Faulkes Telescope South (FTS) in Australia ($i'$). In addition, OGLE and MOA increased the observational cadence with their survey telescopes. MOA also utilized the 0.61m B\&C telescope at MJUO with $I$-band filter to follow up the event. On May 21 13:30 UT ($\mathrm{JD}'=6069.06$), MOA reported the detection of an anomaly as a result of its high cadence observations. The event peaked at around May 21 13:20 ($\mathrm{JD}'=6069.06$) with the maximum magnification of more than 600, around which the event was well covered by several telescopes in New Zealand and Australia. As a result, a quick look light curve clearly showed an anomalous asymmetric feature around the peak, meaning that the lens star hosted a companion. Prompt analyses of the light curve by several teams indicated that this was a firm planetary event having the planet-star mass ratio of $\sim 10^{-3}$. We summarize these photometric observations in Table \ref{tbl:obslog}, and show the observed light curve in Figure \ref{fig:lc_all}. 
We note that data obtained by Possum, Farm Cove, and FTN are not used for the analysis due to the following reasons; data from Farm Cove do not constrain the model because they consist of just two short epochs, and data from Possum and FTN are of relatively low quality.

In addition to the light curve, a high-resolution spectrum of the source star was obtained by \citet{2013A&A...549A.147B} with VLT/UVES on 2012 May 19 ($\mathrm{JD}'=6067$), when the source was magnified by a factor of $\sim$50. They reported that the source star was a metal-poor G dwarf with $T_\mathrm{eff}=5907 \pm 89$ K, $\log g = 4.40 \pm 0.10$, and [Fe/H]$=-0.66 \pm0.07$.

After the event was over, all the observed photometric data were carefully re-reduced by their respective teams. For the re-reductions, OGLE and MOA used their customized pipelines as described in \citet{2015AcA....65....1U} and \citet{2001MNRAS.327..868B}, respectively; $\mu$FUN used the DoPHOT package \citep{1993PASP..105.1342S} for the CTIO 1.3m $H$-band data and a variant of the PySIS package \citep{2009MNRAS.397.2099A} for the others; and RoboNET used the DanDIA package \citep{2008MNRAS.386L..77B}. All the pipelines except for DoPHOT apply the Difference Image Analysis (DIA) method for photometry in order to achieve good photometric precisions in the star-crowding field. 
Note that because a bright neighboring star was located 2\farcs2 away from the target star, the photometry was carefully done to minimize systematics (e.g., by masking the bright star).
The time for each data point was converted to heliocentric julian day (HJD) based on UT.

\begin{deluxetable*}{ccccc}
\tablecaption{Summary of Observations
\label{tbl:obslog}}
\tablewidth{14cm}
\tabletypesize{\small}
\tablehead{
Telescope & Diameter (m) & Filter & $N_\mathrm{use}$ $^\mathrm{a}$/$N_\mathrm{obs}$ $^\mathrm{b}$ 
}
\startdata
The OGLE collaboration &&&\\
OGLE & 1.3  & $I$ & 700/751\\[3pt]
The MOA collaboration &&&\\
MOA-II & 1.8 & $R+I$ & 5449/5947 \\
B\&C & 0.61 & $I$ & 385/439\\[3pt]
The $\mu$FUN collaboration &&&\\
CTIO 1.3m & 1.3 & $V$ & 12/12\\
CTIO 1.3m & 1.3 & $I$ & 30/31\\
CTIO 1.3m & 1.3 & $H$ & 110/140\\
PEST & 0.30 & clear & 163/172\\
Auckland & 0.40 & $R$ & 44/46 \\
Possum & 0.36 & $R$ & 0/21 \\
Farm Cove & 0.36 & clear & 0/23\\[3pt]
The RoboNET collaboration &&&\\
FTS & 2.0 & $i'$ & 195/218\\
LT & 2.0 & $i'$ & 20/23\\
FTN & 2.0 & $i'$ & 0/10
\enddata
\tablenotetext{a}{\ The number of data points used in the analysis.}
\tablenotetext{b}{\ The number of observed data points.}
\end{deluxetable*}

\subsection{Subaru AO Imaging}

We conducted high-resolution imaging of the event field by using the 8.2m {\it Subaru} telescope equipped with the AO instrument AO188 and Infrared Camera and Spectrograph \citep[IRCS,][]{2000SPIE.4008.1056K} at 6:06--7:03 UT on 2012 July 28 ($\mathrm{JD}'=6137.8$), when the source star was still magnified by a factor of 1.47. We used the ``high-resolution" mode of IRCS, which provides a pixel scale of 20.6mas pixel$^{-1}$ and a field of view (FOV) of 21\farcs1 $\times$ 21\farcs1.
We used the natural guide star (NGS) mode for AO. As an NGS, we selected a  bright neighboring star having $R$=11 and being separated by 26$''$ from the source star.
The field was observed through $J$-, $H$-, and $K'$-band filters, each with 30 s exposure $\times$ 15 times at five dithering points within 2$''$ (100 pixel) square. The airmass toward the target field was 1.74--1.54 during the observations. The natural seeing was $\sim$0\farcs5--0\farcs6,  and the AO-worked seeing was 0\farcs23--0\farcs30, 0\farcs19--0\farcs24, and 0\farcs17--0\farcs25 for $J$, $H$, and $K'$, respectively.

All the $J$-, $H$-, and $K'$-band AO images were dark-subtracted and flat-fielded in a standard manner. For the flat fielding, we used sky flat images obtained during evening twilight on the observation night. All the images in each band were combined in average, after the image positions were aligned and sky levels were subtracted. The image-overlapping region was trimmed for further analyses, which resulted in the effective FOV of 18$'' \times$ 18$''$.

\section{Light Curve Analysis}
\label{sec:lc-anal}

\subsection{Model Description}
\label{sec:lc-model}
Because the light curve of the event clearly shows an asymmetric feature around the peak that cannot be explained by a standard single-lens microlensing model, we introduce a binary-lens microlensing model assuming that the lens consists of two objects. The standard binary-lens model can be described with seven basic parameters: the time of the closest source approach to the binary centroid, $t_0$; the Einstein radius ($R_\mathrm{E}$) crossing time, $t_\mathrm{E}$; the minimum impact parameter, $u_0$; the mass ratio of the binary components, $q$; the projected separation of the binary components in units of $R_\mathrm{E}$, $s$; the angle between the source star trajectory and the binary-lens axis, $\alpha$; and the angular source radius in units of $\theta_\mathrm{E}$, $\rho$. In addition, the model requires two instrument-dependent parameters of the unmagnified source flux $F_\mathrm{S}$ and the blended flux $F_\mathrm{B}$. We model the intensity distribution on the surface of the source star with a linear limb-darkening raw of  $I(\theta) = I(0)[1-u_X(1-\cos \theta)]$, where $\theta$ is the angle between the normal to the stellar surface and the line of sight, $I(\theta)$ is the stellar intensity as a function of $\theta$, and $u_X$ is a coefficient for filter $X$. For $u_X$, we adopt the theoretical values of \citet{2013A&A...552A..16C} for a G dwarf with the temperature of 5900 K and log $g$=4.5, namely, $u_V$=0.656, $u_R$=0.572, $u_{R+I}$=0.528, $u_{i'}$=0.504, $u_I$=0.483, and $u_H$=0.290. The $u_{R+I}$ value is calculated as the mean of $u_R$ and $u_I$. For clear filter, we adopt $u_R$ as an approximation. 
The model calculation is done by using a customized code developed by MOA, which is based on the image-centered ray-shooting method \citep{1996ApJ...472..660B,2010ApJ...716.1408B}.

\subsection{Error Normalization}
The initially calculated flux uncertainties in the light curves are normalized as follows.
First, all the light curves are simultaneously fit with the binary-lens models by the Markov Chain Monte Carlo (MCMC) method following \citet{2010ApJ...710.1641S}. Then the flux uncertainties in each light curve are scaled by a single factor so that the $\chi^2$ per degrees of freedom, $\chi_\mathrm{red}^2$, for each light curve becomes unity. This process is iterated until the best-fit light curve model becomes converged. 
Next, after excluding 4$\sigma$ outliers, we renormalize the flux uncertainties for each light curve using the following formula,
\begin{eqnarray}
\sigma_i' = k \sqrt{\sigma_i^2 + e_\mathrm{min}^2},
\end{eqnarray}
where $\sigma_i$ is the initial error bar of the $i$th data point in magnitude, and $k$ and $e_\mathrm{min}$ are coefficients for each data set. Here, the term $e_\mathrm{min}$ represents systematic errors that dominate when the stellar brightness significantly increases.  We adjust the $k$ and $e_\mathrm{min}$ values so that the cumulative $\chi^2$ distribution sorted by magnitude is close to linear, and $\chi_\mathrm{red}^2$ becomes unity.
Finally, all the normalized light curves are fit again and the flux uncertainties are rescaled by a single factor so that $\chi_\mathrm{red}^2$ for all data becomes unity.

\subsection{Best-fit Models}
\label{sec:MCMC}
\begin{figure}
\begin{center}
\includegraphics[width=8cm]{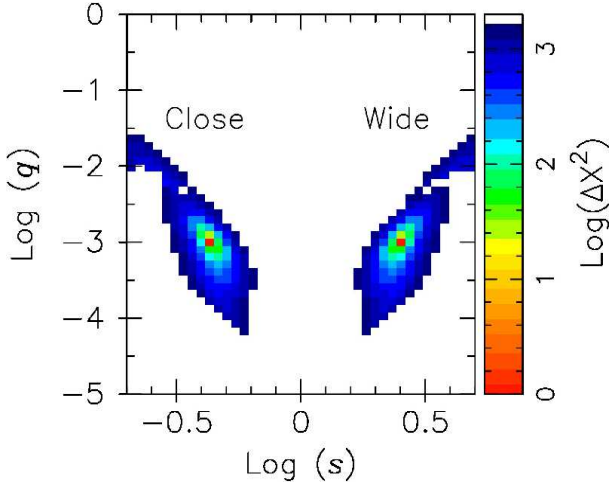}
\caption{A $\chi^2$ map in the $\log s$ vs. $\log q$ plane. 
}
\label{fig:chi2_map}
\end{center}
\end{figure}

In modeling binary microlensing light curves, several models are often degenerate.
In particular for central-caustic crossing and approaching events, a severe degeneracy between $s$ and $s^{-1}$ often occurs because the central-caustic pattern created by a binary lens with $s$ and that with $s^{-1}$ can be very similar  \citep[wide-close degeneracy;][]{1999A&A...349..108D}.
To check for this type of degeneracy and possible other degeneracies, we calculate an $\chi^2$ map in the $\log q$ and $\log s$ plane by dividing $\log q$ ([-5, 0]) and $\log s$ ([-0.7, 0.7]) into 50 $\times$ 40 grids and fitting the light curve while fixing $\log q$ and $\log s$ at each section. These $\log q$ and $\log s$ ranges are chosen such that the usual sensitivity region of microlensing is well covered. The calculated $\chi^2$ map is shown in Figure \ref{fig:chi2_map}. The map shows that there are two local minima at ($\log s$, $\log q$) $\sim$ (-0.4, -3) and (0.4, -3), indicating that the wide-close degeneracy clearly exists.
On the other hand, there is no other local minimum in the map, and the $q$ value at the two local minima of $\sim10^{-3}$ is well within the planetary range ($q<10^{-2}$ for G- and later-type dwarf hosts) with $\Delta \chi^2 > 1000$ over the models of $q>10^{-2}$. Therefore, the planetary nature of this event is quite robust. Note that one can wonder whether there are any other degenerated binary (large-mass ratio) models outside the searched $\log s$ range. However, a very close binary with $\log s < -0.7$ would cause a higher-order effect in the light curve, due to its orbital motion, which we do not detect. In addition, we also check for very wide binary models with $0.7 < \log s < 2$, but find no comparable models with the planetary one.

To properly derive the best-fit model parameters and their uncertainties for the wide and close models, we re-run the MCMC program by starting with the parameter values at each local minimum and letting all parameters be free. For each model, we conduct 40 independent MCMC runs with 3200 steps each, and create posterior probability distributions of the parameters from the merged $\sim$10$^5$ steps. The best-fit value and its 1$\sigma$ lower (upper) uncertainties for each parameter are calculated as the median and 15.9 percentile (84.1 percentile) values of the posterior probability distribution, respectively. The best-fit light curve and caustic models are shown in Figure \ref{fig:lc_all}, while the resultant parameter values, their uncertainties, and the minimum-$\chi^2$ values for the wide and close models are listed in Table \ref{tbl:MCMC}.  The difference of the minimum-$\chi^2$ values of the two models is only 0.6, meaning that these two models are indistinguishable. Note that all the parameter values except for $s$ are almost identical between the two models, and we proceed with further analyses of only the close model as representative of the two, unless $s$ is relevant.

%%

%% MCMC results
\begin{deluxetable*}{lccc}[t]
\tablewidth{14cm}
\tablecaption{Results of the MCMC Analysis. \label{tbl:MCMC}}
\tablehead{
Parameter & Units & Close & Wide
}
\startdata
$t_0$ & HJD-2450000 & 6069.02790 $\pm$ 0.00026 & 6069.02811 $\pm$ 0.00027  \\
$t_\mathrm{E}$ & days & 77.5 $\pm$ 2.2 & 77.7 $\pm$ 2.1 \\
$u_0$ & $10^{-3}$ & 1.405 $\pm$ 0.040  & 1.403 $\pm$ 0.039 \\
$q$ & $10^{-3}$ & 1.086 $\pm$ 0.040 & 1.085 $\pm$ 0.039 \\
$s$ & $\theta_\mathrm{E}$ & 0.4134 $\pm$ 0.0032  & 2.425 $\pm$ 0.019\\
$\alpha$ & rad & 5.7843 $\pm$ 0.0013 & 5.7841 $\pm$ 0.0013\\
$\rho$ & $10^{-4}$ $\theta_\mathrm{E}$ & 3.27 $^{+0.24}_{-0.26}$ & 3.27 $^{+0.24}_{-0.26}$  \\
\hline
$\chi^2_\mathrm{min}$ & & 7068.6 & 7069.2
\enddata
\end{deluxetable*}

\subsection{Searching for the Microlens Parallax Effect}
\label{sec:piE_limit}

In order to know the absolute masses and projected separation of the planetary system instead of $q$ and $s$, one needs to derive the total mass of the planetary system, $m_\mathrm{L}$, and the distance from the Earth to the system, $D_\mathrm{L}$. One way to do this is to detect the microlens parallax effect in the light curve, which is seen as a slightly asymmetric distortion over the non-parallax light curve. The distortion depends on $m_\mathrm{L}$, $D_\mathrm{L}$, the vectors of the Earth's orbital acceleration, and the lens-source relative proper motion \citep{1992ApJ...392..442G}. From this effect, one can obtain an additional parameter $\pi_\mathrm{E}$, which is defined as the ratio of 1 AU to the projected Einstein radius onto the Earth's plane \citep{2000ApJ...542..785G}.  From $\pi_\mathrm{E}$ and the angular Einstein radius, $\theta_\mathrm{E}$, which will be measured in Section \ref{sec:CMD}, one can derive $m_\mathrm{L}$ and $D_\mathrm{L}$ as
\begin{eqnarray}
m_\mathrm{L} &=& \frac{\theta_\mathrm{E}}{\kappa \pi_\mathrm{E}}\\
D_\mathrm{L} &=& \frac{\mathrm{AU}}{\pi_\mathrm{E} \theta_\mathrm{E} + 1/D_\mathrm{S}},
\end{eqnarray}
where $\kappa$ $\equiv 4G/c^2$ $\simeq 8.144$ $\mathrm{mas}~\mathrm{M}_\odot^{-1}$ and $D_\mathrm{S}$ is the source distance. Here $G$ is the gravitational constant and $c$ is the speed of light.

To search for the microlens parallax effect, we fit the light curve by freeing two additional parameters of $\pi_\mathrm{E,E}$ and $\pi_\mathrm{E,N}$, which are the east and north components of {\boldmath $\pi_\mathrm{E}$}, respectively, where {\boldmath $\pi_\mathrm{E}$} is a vector whose length is $\pi_\mathrm{E}$ and direction is the same as the lens direction relative to the source. 

As a result, the parallax model fit gives the best-fit values of $\pi_\mathrm{E,E}$ =-0.024 $\pm$ 0.058 and $\pi_\mathrm{E,N}$=0.62 $\pm$ 0.17 with the $\chi^2$ of 7054.1, which has the improvement of 14.5 compared with that for the non-parallax model. Statistically speaking, this is a marginal (3.8$\sigma$) detection of the parallax signal, however, we consider it suspect because of the following reasons.

\begin{figure}
\begin{center}
\includegraphics[width=8.5cm]{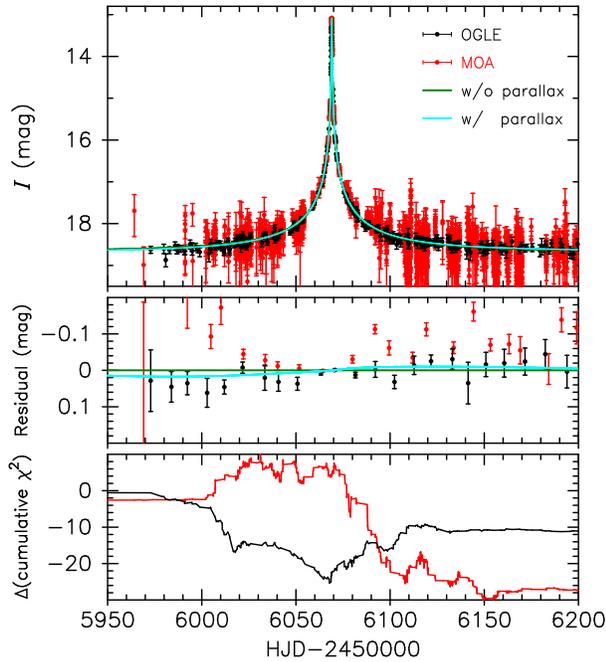}
\caption{
(Top) The OGLE (black) and MOA-II (red) light curves along with the best-fit models with (green) and without (cyan) the parallax effect. (Middle) Residuals from the non-parallax best-fit model. The data points are binned in 10 days for clarity. (Bottom) Difference of the cumulative $\chi^2$ between the parallax and non-parallax models for the OGLE (black) and MOA-II (red) data. A negative value means a preference for the parallax model. Note that, although this plot indicates that the total $\chi^2$ improvement from these two data sets is $\sim$38, that from all data sets is 14.5, as described in text.
}
\label{fig:para_vs_nonpara}
\end{center}
\end{figure}

The difference between the best-fit parallax and non-parallax models emerges around the wings of the light curve, where the MOA-II and OGLE data are dominant. 
In Figure \ref{fig:para_vs_nonpara}, we show the MOA-II and OGLE light curves along with the best-fit parallax (green) and non-parallax (cyan) models in the top panel, the residuals of the observed light curves against the non-parallax model in the middle panel, and the difference of cumulative $\chi^2$ between the two models in the bottom panel. The residual plot shows that the difference of the two models is quite small compared to the error bars of the data points.
In such a case, systematics in a small number of data points can lead to a false positive detection. In fact, as indicated by the $\Delta$cumulative-$\chi^2$ plot, the major $\chi^2$ improvement comes from the MOA-II data in the limited range of 6070 $\lesssim$ HJD-2450000 $\lesssim$ 6110, although the model difference emerges more widely over several hundred days. Moreover, the $\Delta$cumulative-$\chi^2$ plots for the MOA-II and OGLE data are anti-correlated rather than correlated, indicating that the detected parallax signal is suspicious. Therefore, the detected signal is probably a false positive, and we only calculate an upper limit on $\pi_\mathrm{E} < 1.53$ (3 $\sigma$).
We note that other effects that can mimic the parallax signal, such as the orbital motion of the source and/or lens systems, can also be rejected for the same reasons.

\section{Properties of the Source Star}
\label{sec:CMD}

The $\theta_E$ value can be derived by dividing $\rho$ by the angular radius of the source star $\theta_*$, which can be estimated from the intrinsic color and magnitude of the source star. 
For most microlensing events, no spectroscopic information of the source star is available, due to the intrinsic faintness of the star. In this case, the intrinsic color and magnitude of the source star are estimated by using red clump stars in the Galactic bulge as a standard candle, assuming that the dust extinction for the source star is the same as that for those red clump stars. We first derive $\theta_*$ using this ``standard'' method in Section \ref{sec:theta*_standard}. On the other hand, in the current case, a high-resolution spectrum of the source star was obtained during the high-magnification state of the event by \citet{2013A&A...549A.147B}, as part of a systematic elemental-abundance study of the bulge stars. With this spectral information, the source's intrinsic color and magnitude can be derived with a minimum assumption about the extinction for the source star. We show how the spectral information improves the $\theta_*$ estimation in Section \ref{sec:theta*_w_spec}. In addition, we measure the distance to the source star by combining the photometric and spectroscopic information in Section \ref{sec:D_S}.

\subsection{Angular-radius Estimation Using a Standard Method}
\label{sec:theta*_standard}

We first derive the apparent $I$- and $V$-band magnitudes of the source star, $I_\mathrm{S}$ and $V_\mathrm{S}$, respectively, from the light-curve fitting.
We derive $I_\mathrm{S}$ from the two light-curve data sets obtained by the OGLE and CTIO 1.3m telescopes. Calibrating the instrumental magnitude to the standard (Landolt) one via the OGLE-III catalog \citep{2011AcA....61...83S}, we obtain $I_\mathrm{S} = 20.147 \pm 0.031$ and  $I_\mathrm{S} = 20.082 \pm 0.031$ from the respective data sets. 
By taking the mean of the two values, we obtain a final value of $I_\mathrm{S}$ = 20.115 ± 0.031. Note that the uncertainty of $I_\mathrm{S}$ is conserved in this calculation because this uncertainty is dominated by the light-curve model uncertainty and therefore the two $I_\mathrm{S}$ values are correlated.
$V_\mathrm{S}$ is derived from the $V$-band light curve obtained by the CTIO 1.3m telescope. In the same way as for the I band, we derive the calibrated source magnitude of $V_\mathrm{S}=21.595 \pm 0.032$. Then, from  $I_\mathrm{S}$ and $V_\mathrm{S}$,  we obtain the apparent source color of $(V-I)_\mathrm{S}$ = 1.480 $\pm$ 0.032. 

\begin{figure}
\begin{center}
\includegraphics[width=8cm]{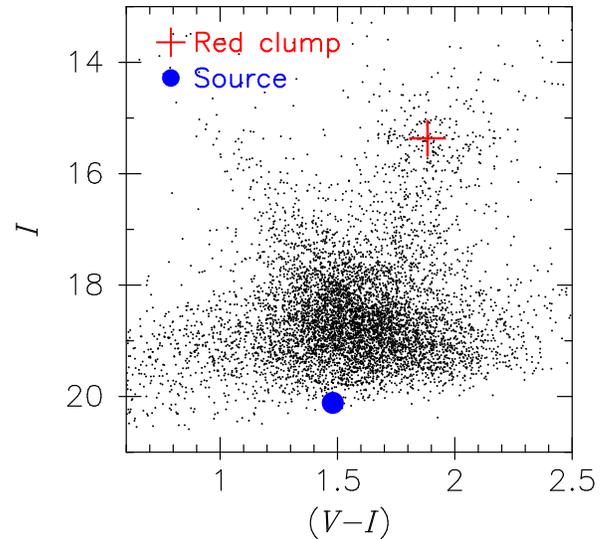}
\caption{
Color-magnitude diagram (CMD) toward the event field ($\sim$2\farcm8 $\times$ 2\farcm8 area) created from the OGLE-III catalog. The centroid of red clump giants and the source position are indicated by the red cross and blue circle, respectively.
}
\label{fig:CMD}
\end{center}
\end{figure}

Next, we estimate the extinction and reddening for the source star using red clump stars in the Galactic bulge. In Figure \ref{fig:CMD}, we plot the color magnitude diagram (CMD) toward the event coordinate ($\sim$2\farcm8 $\times$ 2\farcm8 area) created from  the OGLE-III catalog, along with the measured $I_\mathrm{S}$ and ($V-I$)$_\mathrm{S}$. 
The source position is largely consistent with, but a bit fainter than, the region of MS stars in the Galactic bulge, which is consistent with the fact that the source is located behind the center of the bulge along the line of sight (see Section \ref{sec:D_S}).
The centroid of the red clump stars on the CMD is measured as ($V-I$)$_\mathrm{RC}=1.883\pm0.013$ and $I_\mathrm{RC}=15.361\pm0.015$, respectively. These values are then compared to the intrinsic color and magnitude of the red clump stars of ($V-I$)$_\mathrm{RC,0}=1.06\pm0.06$ \citep{2011A&A...533A.134B} and $I_\mathrm{RC, 0}=14.34\pm0.04$ \citep{2013ApJ...769...88N}, leading to the reddening and extinction magnitudes for the red clump stars of $E(V-I)=0.82\pm0.06$ and $A_I=1.02\pm0.04$, respectively. 
Assuming that this reddening and extinction can also be applied for the source star, we derive the intrinsic color and magnitude of the source star as ($V-I$)$_\mathrm{S, 0}=0.66\pm0.07$ and $I_\mathrm{S, 0}=19.10\pm0.05$, respectively.

Finally, we estimate the angular radius of the source star $\theta_*$ from $I_\mathrm{S,0}$ and ($V-I$)$_\mathrm{S,0}$ using the following equation: 
\begin{eqnarray}
\label{eq:VI_to_theta_s}
\log \theta_\mathrm{LD} = 0.5014 +  0.4197 (V-I) - 0.2 I,
\end{eqnarray}
 where $\theta_\mathrm{LD}$ is the limb-darkened stellar angular diameter. This linear equation is derived from a subset of the interferometrically measured stellar radii presented in \citet{2014AJ....147...47B}, restricting stars with 3900 K $< T_\mathrm{eff}<$ 7000 K to improve the fit for FGK stars. This gives $\theta_* \equiv \theta_\mathrm{LD}/2 = 0.454 \pm 0.033$ $\mu$as.

The derived source properties are summarized in Table \ref{tbl:source_properties}.

\begin{deluxetable}{ccc}
\tablewidth{8cm}
\tablecaption{Properties of the Source Star \label{tbl:source_properties}}
\tabletypesize{\small}
\tablehead{
Parameter & Value &Value\\
& (w/o spec. info) & (w/ spec. info) 
}
\startdata
$T_\mathrm{eff}$ (K) & --- & 5907 $\pm$ 89 $^\mathrm{a}$\\
\ [Fe/H] & --- & -0.66 $\pm$ 0.07 $^\mathrm{a}$\\
$\log g$ (cgs) & --- & 4.40 $\pm$ 0.10 $^\mathrm{a}$\\
\hline
$V_\mathrm{S}$ & 21.595 $\pm$ 0.032 & ---\\
$I_\mathrm{S}$ & 20.115 $\pm$ 0.031 & ---\\
$H_\mathrm{S}$ & 18.569 $\pm$ 0.032 & ---\\
\hline
$(V-I)_\mathrm{S,0}$ & 0.66 $\pm$ 0.07 & 0.660 $\pm$ 0.027\\
$(V-H)_\mathrm{S,0}$ & --- & 1.41 $\pm$ 0.06\\
$V_\mathrm{S,0}$ & 19.76 $\pm$ 0.08 & 19.73 $\pm$ 0.07\\
$I_\mathrm{S,0}$ & 19.10 $\pm$ 0.05 & 19.14 $\pm$ 0.08\\
$H_\mathrm{S,0}$ & --- & 18.314 $\pm$ 0.041\\
\hline
$E(V-I)$ & 0.82 $\pm$ 0.06 & 0.820 $\pm$ 0.042\\
$E(J-K_\mathrm{s})$ & 0.244 $\pm$ 0.109 $^\mathrm{b}$ & 0.266 $\pm$ 0.019\\
$A_V$ & 1.84 $\pm$ 0.07 & 1.80 $\pm$ 0.09\\
$A_I$ & 1.02 $\pm$ 0.04 & 0.98 $\pm$ 0.08\\
$A_J$& --- & 0.423 $\pm$ 0.024\\
$A_H$& --- & 0.255 $\pm$ 0.029\\
$A_{K_\mathrm{s}}$& --- & 0.157 $\pm$ 0.017\\
\hline
$\theta_*$ [$\mu$as] ($I$, $V-I$) & 0.454 $\pm$ 0.033&  \rev{{\bf0.446 $\pm$ 0.023} $^\mathrm{c}$}\\
$\theta_*$ [$\mu$as] ($H$, $V-H$) & --- & 0.444 $\pm$ 0.014\\
\hline
$M_V$ & --- & 4.99 $^{+0.26}_{-0.14}$\\
$M_I$ & --- & 4.28 $^{+0.25}_{-0.19}$\\
$M_H$ & --- & 3.54 $^{+0.24}_{-0.16}$\\
$t$ (Gyr) & --- & 13 $^{+0}_{-5}$\\
$D_\mathrm{S}$ (kpc) & --- & 9.1 $^{+0.9}_{-1.1}$
\enddata
\tablenotetext{a}{\ From \citet{2013A&A...549A.147B}.}
\tablenotetext{b}{\ From the BEAM calculator \citep{2012A&A...543A..13G}.}
\tablenotetext{c}{\rev{\ The value adopted for the rest of analyses.}}
\end{deluxetable}

\subsection{Angular-radius Estimation with Spectral Information}
\label{sec:theta*_w_spec}

\subsubsection{From $(V-I)$ and $I$}
\label{sec:from_VI}
The spectroscopically measured effective temperature $T_\mathrm{eff} = 5907 \pm 89$ K and metallicity  [Fe/H] = -0.66 $\pm$ 0.07 \citep{2013A&A...549A.147B} of the source star can be directly converted to the source's intrinsic color. 
Using the color-metallicity-temperature relation of \citet{2010A&A...512A..54C}, we derive ($V-I$)$_\mathrm{S,0} = 0.660 \pm 0.027$. This value is in good agreement with the photometrically derived ($V-I$)$_\mathrm{S,0}$ but has a smaller uncertainty by a factor of $\sim$2.5. Subtracting this value from the apparent source color of ($V-I$)$_\mathrm{S}=1.480 \pm 0.032$, we derive the reddening magnitude up to the source star of $E(V-I) = 0.820 \pm 0.042$. Then, assuming that the following extinction law 
\begin{eqnarray}
\label{eq:A_I}
A_I = &&1.217 \times \nonumber\\
&&E(V-I)  [1+1.126 \times (R_{JKVI}-0.3433)]
\end{eqnarray}
 \citep{2013ApJ...769...88N} applies along the line of sight, and adopting $R_{JKVI} \equiv E(J-K_\mathrm{s})/E(V-I) = 0.3249$ for the event coordinate from the online Extinction Calculator\footnote{http://ogle.astrouw.edu.pl} \citep{2013ApJ...769...88N}, we obtain $A_I = 0.98 \pm 0.08$ (the error includes the uncertainty of Equation (\ref{eq:A_I}), for which we adopt 0.06 mag). Consequently, we derive the intrinsic source magnitude of $I_\mathrm{S,0} = I_\mathrm{S} - A_I = 19.14 \pm 0.08$. This value is also consistent with the previous estimation but has a bit larger uncertainty, which mainly comes from the uncertainty of the estimation of $A_I$. Note, however, that this $A_I$ estimation is not based on any assumption about the absolute extinction or reddening, but on the  extinction law of Equation (\ref{eq:A_I}). The source angular radius is then calculated by using Equation (\ref{eq:VI_to_theta_s}) as 
 \begin{eqnarray}
 \theta_* = 0.446 \pm 0.023\ \mu\mathrm{as},
 \end{eqnarray}
 whose uncertainty is 1.4 times smaller than the previous one.
 \rev{
 Using this $\theta_*$ and $\rho$, we derive the angular Einstein  radius of 
\begin{eqnarray}
\theta_\mathrm{E} = 1.36 ^{+0.14}_{-0.12}\ \mathrm{mas}.
\end{eqnarray}
 We also derive the geocentric source-lens relative proper motion from $\theta_\mathrm{E}$ and  $t_\mathrm{E}$ as 
\begin{eqnarray}
\mu_\mathrm{geo} = 6.4 ^{+0.6}_{-0.5} \ \mathrm{mas\ yr^{-1}}.
\end{eqnarray}
}

\subsubsection{From $(V-H)$ and $H$}
\label{sec:from_VH}
We recalculate $\theta_*$ from $(V-H)$ and $H$ of the source star in order to check for the robustness of the previous estimation. 
In general, one can obtain a better estimation of $\theta_*$ from $(V-H)$ and $H$ rather than from $(V-I)$ and $I$ \citep[c.f., Table 4 of][]{2004A&A...426..297K}.

In the same way as for ($V-I$)$_\mathrm{S,0}$, we calculate the intrinsic ($V-H$) source color of ($V-H$)$_\mathrm{S,0} = 1.41 \pm 0.06$, where $H$ is in the 2MASS system.
Then, the intrinsic $H$-band source magnitude $H_\mathrm{S,0}$ is estimated as follows. 
First, the apparent $H$-band source magnitude is measured from the microlensing fit to the $H$-band light curve obtained by the CTIO 1.3m telescope, yielding $H_\mathrm{S} = 18.569 \pm 0.032$ in the 2MASS system. Next, the $H$-band extinction $A_H$ is estimated by combining the $(J-K_\mathrm{s})$ reddening of $E(J-K_\mathrm{s})=R_{JKVI} \times E(V-I) = 0.266 \pm 0.019$ (adopting 0.01 for the uncertainty of $R_{JKVI}$) and two extinction coefficients of $A_J/A_{K_\mathrm{s}}$ and $A_H/A_{K_\mathrm{s}}$. We adopt $A_J/A_{K_\mathrm{s}}=2.70 \pm 0.15$ and $A_H/A_{K_\mathrm{s}}=1.63 \pm 0.05$, which are the mean values for the Galactic bulge estimated by \citet{2013A&A...550A..42C}, but have conservative uncertainties taking into account the non-uniformity of these coefficients toward the Galactic bulge.
 We solve these equations for $A_H$ and derive $A_H=0.255 \pm 0.029$. As a by-product, we also obtain $A_J=0.423 \pm 0.024$ and $A_{K_\mathrm{s}}=0.157 \pm 0.017$, which will be used in Section \ref{sec:excess_flux}. Finally, we derive $H_\mathrm{S,0} = H_\mathrm{S} - A_H = 18.314 \pm 0.041$.

The derived ($V-H$)$_\mathrm{S,0}$ and $H_\mathrm{S,0}$ values are then converted to $\theta_*$ using the following equation:
\begin{eqnarray}
\label{eq:VH_to_theta_s}
\log \theta_\mathrm{LD} = 0.53598 &+& 0.07427 (V-H) \nonumber\\
 &+& 0.04511 \mathrm{[Fe/H]} - 0.2 H,
\end{eqnarray}
where $H$ is in the Johnson magnitude system. This equation is derived in the same way as for Equation (\ref{eq:VI_to_theta_s}), but includes the metallicity term because a small metallicity dependence appears in this relation. We adopt [Fe/H]=-0.66 $\pm$ 0.07 from \citet{2013A&A...549A.147B}. We convert ($V-H$) in the 2MASS magnitude to that in the Johnson system via the transformation of ($V-H_\mathrm{Johnson}$) $= 0.28302 \times 10^{-2}  + 1.0021 (V-H_\mathrm{2MASS}) + 0.36618 \times 10^{-2} (V-H_\mathrm{2MASS})^2 - 0.17906 \times 10^{-2} (V-H_\mathrm{2MASS})^3 + 0.16113 \times 10^{-3} (V-H_\mathrm{2MASS})^4$, which is derived by combining Equation (B5) and (B8) from \citet{2001AJ....121.2851C} and a color-color relation of Table A5 from \citet{1995ApJS..101..117K}.  As a result, we derive 
$\theta_* = 0.444 \pm 0.014\ \mu\mathrm{as}$,
which is quite consistent with the previous estimations, and even has the smallest uncertainty. 
However, this value is derived partly by relying on the two extinction coefficients of $A_J/A_{K_\mathrm{s}}$ and $A_H/A_{K_\mathrm{s}}$, which can vary depending on the line of sight \citep{2013A&A...550A..42C} although we adopt the mean values for the galactic bulge. That could produce some systematics that are difficult to asses. Therefore, we adopt the $\theta_*$ value derived in the previous section for the rest of analyses to be conservative.

\subsection{Distance to the Source Star}
\label{sec:D_S}

The distance to the source star $D_\mathrm{S}$ can be measured by estimating the absolute magnitude of the source star in addition to the intrinsic one.
To this end, we use the isochrone models of PARSEC version 1.2S \citep{2012MNRAS.427..127B}. Through the web interface CMD 2.7,\footnote{http://stev.oapd.inaf.it/cgi-bin/cmd\_2.7} we obtain a grid of isochrone tables for the stellar age in the range of 8.0 $< \log t \mathrm{[yr]} <$ 10.13  with a step size of $\log t$=0.02, and for the metallicity in the range of -0.88 $<$ [M/H] $<$ -0.45 with a step size of 0.035. For each table, we further grid the table interpolatively by $\log T_\mathrm{eff}$ with a step size of 0.001. Then, for each grid, we calculate the following $\chi^2$ value
\begin{eqnarray}
\chi^2 = \frac{(T_\mathrm{eff} - T)^2}{\sigma_{T_\mathrm{eff}}^2}
+ \frac{(\mathrm{[Fe/H]} - M)^2}{\sigma_\mathrm{[Fe/H]}^2}
+ \frac{(\log g - G)^2}{\sigma_{\log g}^2},
\end{eqnarray}
where $T$, $M$, and $G$ are the model temperature, metallicity, and surface gravity, respectively. We then find the best-fit values of the absolute $V$, $I$, and $H$ magnitudes $M_V$, $M_I$, and $M_H$, as well as the stellar age $t$, by minimizing the $\chi^2$ value. In addition, 1$\sigma$ uncertainties of these parameters are calculated by searching for the region where $\Delta \chi^2=1$.
The resultant values and uncertainties are $M_V = 4.99 ^{+0.26}_{-0.14}$, $M_I =  4.28 ^{+0.25}_{-0.19}$, $M_H = 3.54 ^{+0.24}_{-0.16}$, and $t = 13 ^{+0}_{-5}$ Gyr. We note that \citet{2013A&A...549A.147B} also estimated $M_V$, $M_I$ and $t$ by using the Yonsei-Yale isochrone models as $4.89$, $4.23$, and $10.2 ^{+1.8}_{-4.5}$ Gyr, respectively. Our estimations are consistent with theirs.

Combining these absolute magnitudes with the intrinsic ones derived in Section \ref{sec:theta*_w_spec}, we calculate the source distance to be $D_\mathrm{S}$ = 8.8 $^{+0.7}_{-1.0}$, 9.4 $^{+0.9}_{-1.1}$, and 9.0 $^{+0.7}_{-1.0}$ kpc for $V$, $I$, and $H$, respectively. By taking the mean of the three values while conservatively keeping the uncertainties, we obtain a final value of $D_\mathrm{S}$ = 9.1 $^{+0.9}_{-1.1}$ kpc. On the other hand, the distance to the centroid of the red clump stars toward the event coordinate is estimated using the Extinction Calculator of \citet{2013ApJ...769...88N} to be $D_\mathrm{RC} = 7.9 ^{+0.9}_{-0.8}$ kpc, which indicates that the source star is likely located at the far side of the Galactic bulge. This result is in agreement with the fact that the microlensing event rate is higher for far-side source stars because of the higher stellar density per unit solid angle.

\section{Extracting Excess flux}
\label{sec:extracting_excess_flux}

Although the orbital parallax is not detected in the microlensing light curves, the absolute lens mass and distance can be constrained if the flux from the lens star is detected. Although the lens and source stars cannot spatially be resolved at present, the lens flux can be observed as an excess of flux at the source position. To do this, we analyze the high-resolution images obtained with {\it Subaru}/IRCS.

\subsection{Identifying the microlensing target}

\begin{figure}
\begin{center}
\includegraphics[width=8cm]{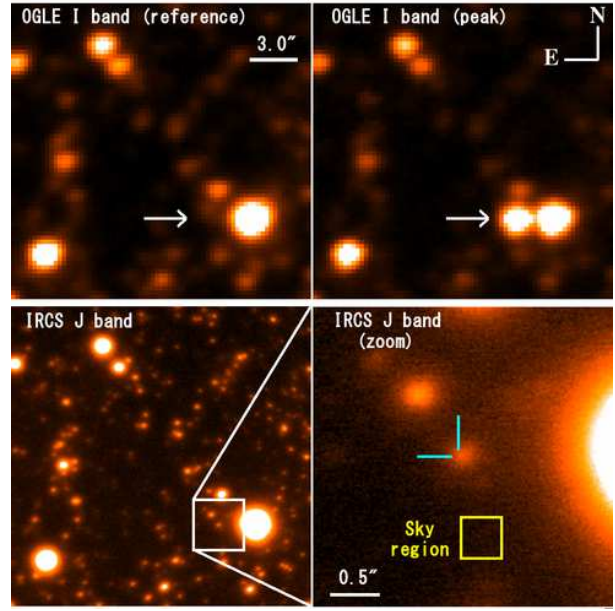}
\caption{
(Upper left) the OGLE $I$-band reference image obtained well before the microlensing event. The location of the source star to be magnified is indicated by an arrow. (Upper right) an OGLE $I$-band image obtained near the event peak. (Lower left) the IRCS $J$-band image. (Lower right) a zoom of the image in the lower-left panel. The calibrated source position is indicated by cyan lines. The region in which the sky value is calculated is shown in yellow box.
}
\label{fig:ds9}
\end{center}
\end{figure}

The first step to extract the excess flux is to identify the microlensing target (source+lens superposition) on the {\it Subaru}/IRCS images. To do so, we use an OGLE-IV $I$-band image obtained near the brightness peak with the source magnification of more than 400 times (upper right panel in Figure \ref{fig:ds9}), which overwhelms any blended fluxes so that the source position can unambiguously be measured.
Because the target position is almost identical to the source position at the time of the {\it Subaru} observation, the measured source position on the OGLE peak image can be used to identify the target on the IRCS image once the instrumental coordinate of the OGLE image is calibrated to that of the IRCS one.
The coordinate calibration is done by constructing a calibration ladder via an OGLE reference image (upper left panel in Figure \ref{fig:ds9}) obtained under a better seeing condition (1\farcs0) compared to the OGLE peak image (1\farcs2).
Specifically, we first calibrate the instrumental coordinate of the OGLE peak image to that of the OGLE reference one using the centroid positions of 18 bright common stars  (first calibration), and then calibrate the instrumental coordinate of the OGLE reference image to that of the IRCS/$J$-band one using the centroid positions of nine well-isolated common stars (second calibration). All the stellar centroids are measured by using the DoPHOT package. The coordinate calibrations are done by using the IRAF\footnote{IRAF is distributed by the National Optical Astronomy Observatories, which are operated by the Association of Universities for Research in Astronomy, Inc., under cooperative agreement with the US National Science Foundation.}   {\it GEOMAP} algorithm with free parameters of $xy$ shifts and rotation for the first calibration, and $xy$ shifts, $xy$ pixel-scale magnifications, and rotation for the second calibration. The rms values for the first and second calibrations are 38 mas for 18 stars and 44 mas for nine stars, respectively, meaning that the total 1$\sigma$ calibration error is 17 mas. We measure the source centroid position on the OGLE peak image with an adopted 1$\sigma$ uncertainty of 0.1 pixel, or 26 mas, which is then calibrated to the coordinate on the IRCS image (marked as cyan lines in Figure \ref{fig:ds9}) with the total uncertainty of  31 mas. On the IRCS image, we find that there is one stellar object close to the source position with a separation of only 44 mas, which is comparable with the measurement uncertainty. We therefore conclude that this object is the microlensing target.

\subsection{Photometry of the microlensing target}

\begin{figure*}
\begin{center}
\includegraphics[width=16cm]{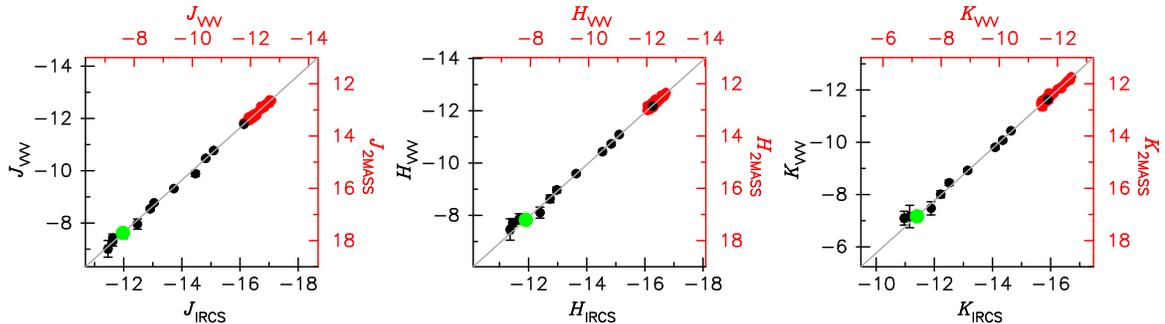}
\caption{
Photometric-calibration ladders  for the $J$ (left), $H$ (middle), and $K_\mathrm{s}$ (right) bands. For all panels, 
the black points indicate the common stars between IRCS and VVV used to calibrate the instrumental magnitudes of IRCS (bottom axis) to that of VVV (left axis), while the red points indicate the common stars between VVV and 2MASS used to calibrate the instrumental VVV magnitudes (top axis) to the 2MASS ones (right axis).
The location of the target object is indicated as green circle.}
\label{fig:calib_ladder}
\end{center}
\end{figure*}

The next step is to measure the brightness of the microlensing target. We perform aperture photometry of the target object on the IRCS $J$-, $H$-, and $K'$-band images, by using a customized code with an aperture radius of 10 pixels (0\farcs20). As a bright star is located at 108 pixels (2\farcs2) west of the target and its point-spread function (PSF) tail spreads toward the target, we carefully select a region to estimate the sky level such that the distance from the centroid of the bright star is the same as for the target and there is no other noticeable flux contamination. The selected region, a box with the size of 20 $\times$ 20 pixels, is indicated in yellow in Figure \ref{fig:ds9}. We calculate a median sky value from this region, and subtract it from the fluxes in the target's aperture. The measured target flux is then calibrated to the 2MASS magnitude system. For this calibration, we construct a photometric calibration ladder using $J$-, $H$-, and $K_\mathrm{s}$-band archive images of VISTA Variables in the Via Lactea \citep[VVV;][]{2010NewA...15..433M}, because there are only three overlapping stars between the IRCS images and the 2MASS catalog. The VVV images are three times finer in pixel scale and four times deeper in limiting magnitude  compared to the 2MASS catalog. We here approximate that the IRCS $J$, $H$, and $K'$ bands are identical to the 2MASS $J$, $H$, and $K_\mathrm{s}$ bands, respectively. On the VVV images, we perform stellar extraction and PSF-fitting photometry on a 2\farcm5 $\times$ 2\farcm5 subarea around the target by using the DoPHOT package.  Among 40--60 common stars with the 2MASS catalog, 30 bright-end stars are used for photometric calibration in order to avoid the effect of blending on the fainter objects. For the {\it Subaru}/IRCS images, we carefully select 12, 11, and 10 calibration stars for $J$, $H$, and $K_\mathrm{s}$, respectively, such that they are well isolated on the IRCS images and also detected on the VVV images. Aperture photometry of these stars is done by the same manner as for the target object, but this time the sky levels are estimated from the stellar-centroid annulus with an inside and outside radii of 70 and 80 pixels (1\farcs4 and 1\farcs6), respectively. The constructed calibration ladders are shown in Figure \ref{fig:calib_ladder}. As a result, we measure the target brightness in $J$, $H$, and $K_\mathrm{s}$ as

\begin{eqnarray}
J_\mathrm{target} &=& 17.704 \pm 0.034 \\
H_\mathrm{target} &=& 17.207 \pm 0.039\\
K_\mathrm{s,target} &=& 17.071 \pm 0.044.
\end{eqnarray}

\subsection{The Excess Flux}
\label{sec:excess_flux}

The final step to extract the excess flux is to subtract the apparent source fluxes from the measured target ones.

Although the apparent $H$-band magnitude of the source star is measured from the light curve fitting as $H_\mathrm{S} = 18.569 \pm 0.032$,  no $J$- or $K_\mathrm{s}$-band light curves were obtained. So, we estimate the apparent $J$- and $K_\mathrm{s}$-band source magnitudes $J_\mathrm{S}$ and $K_\mathrm{s,S}$ as follows.
First, using the color-metallicity-temperature relation of \citet{2010A&A...512A..54C}, we obtain the intrinsic source colors in $(V-J)$ and $(V-K_\mathrm{s})$ as $(V-J)_\mathrm{S,0} = 1.13 \pm 0.05$ and $(V-K_\mathrm{s})_\mathrm{S,0} = 1.48 \pm 0.06$, respectively. Next, substituting these values from $V_\mathrm{S,0}=H_\mathrm{S,0}-(V-H)_\mathrm{S,0}=19.73 \pm 0.07$, we obtain the intrinsic source magnitudes of $J_\mathrm{S,0} = 18.60 \pm 0.09$ and $K_\mathrm{s, S, 0} = 18.25 \pm 0.09$, respectively. Finally, the apparent $J$ and $K_\mathrm{s}$ source magnitudes are derived as $J_\mathrm{S} = 19.02 \pm 0.09$ and $K_\mathrm{s,S} = 18.40 \pm 0.10$ by adding $A_J$ and $A_\mathrm{K_\mathrm{s}}$ to $J_\mathrm{S,0}$ and $K_\mathrm{s, S,0}$, respectively. 

The best-fit light curve models suggest that the source star was still magnified by 1.474 at the time of {\it Subaru} observation, meaning that the source magnitudes in $J$, $H$, and $K_\mathrm{s}$ at the time of the {\it Subaru} observation were $18.60 \pm 0.09$, $18.15 \pm 0.032$, and $17.98 \pm 0.10$, respectively. All of these magnitudes are significantly fainter than the measured target ones, indicating that excess flux at the source position is clearly detected. Subtracting the source fluxes from the target ones, we obtain the $J$-, $H$-, and $K_\mathrm{s}$-band magnitudes of the excess flux:

\begin{eqnarray}
J_\mathrm{excess} &=&18.33 \pm 0.09 \\
H_\mathrm{excess} &=& 17.80 \pm 0.07\\
K_{s,\mathrm{excess}} &=& 17.69 \pm 0.11.
\end{eqnarray}

\subsection{Origin of the Excess Flux}

In principle, there are three possible scenarios for the origin of the detected excess flux: (1) it solely comes from the lens star, (2) it is a combination of fluxes from the host star and from  other astronomical sources (such as  an unrelated chance-alignment star, a companion to the source star, and/or a companion to the lens star), and (3) it entirely comes from other astronomical sources mentioned previously.

In the first case, the lens star is almost certainly an MS star rather than a giant or stellar remnant, given the brightness of the excess flux. In the second case, as we will show in Section \ref{sec:contami}, the amount of the contamination flux from the extra sources is constrained to $\lesssim50$\% of the measured excess flux by the $\theta_\mathrm{E}$ measurement and the upper limit of $\pi_\mathrm{E}$; with this limitation, the lens star is also most likely an MS star. 

In the final case, the lens star must not be an MS star, but rather a stellar remnant such as a white dwarf (WD), a neutron star (NS), or a black hole (BH). This could happen in principle, however we ignore this possibility here because of their relatively low population  in the Galaxy  \citep[MS:WD:NS:BH $\sim$ 1:0.18:0.021:0.0031;][]{2011Natur.473..349S} and the presumably low planetary abundance around them (no planet has yet been detected around a WD or BH). In particular, these objects are created after disruptive stellar evolutions, including radius inflation for all cases and subsequent violent explosions for the case of an NS and BH, which could reduce planetary abundance around them. Secondary planet formation after the evolutions might be possible, however, its efficiency is not yet known.

Therefore, hereafter we simply assume that the lens star is an MS star, and consider only the scenarios (1) and (2). We note that the stellar remnant scenario can be tested by obtaining high-resolution images in the near future when the source and lens stars will be separated enough; we would not see any object at the expected separation if the lens were a remnant.

\section{Physical Parameters of the Lens System}
\label{sec:physical_parameters}

\subsection{No Contamination Assumption}
\label{sec:no_contami}

The absolute physical parameters of the lens system can be derived by combining the lens flux and $\theta_\mathrm{E}$, both of which are independent functions of the host star's mass, $m_\mathrm{host}$, and the distance to the lens system from the Earth, $D_\mathrm{L}$.

\begin{figure}
\begin{center}
\includegraphics[width=8.5cm]{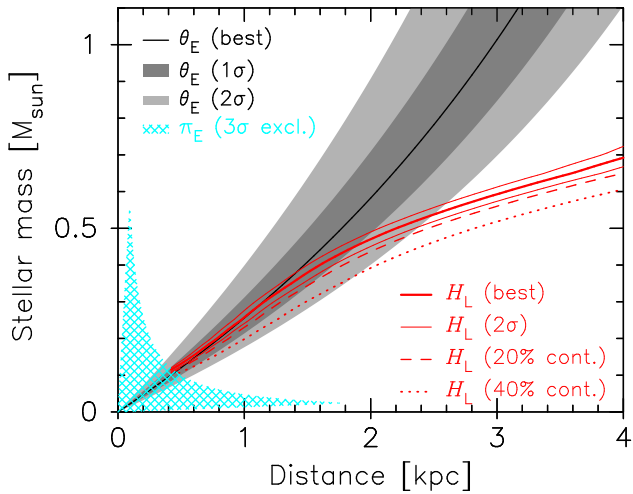}
\caption{
Constraints on the distance and mass of the host star. The black solid line shows a relation from the best-fit $\theta_\mathrm{E}$ value, and the dark gray and light gray regions indicate its 1$\sigma$ and 2$\sigma$ confidence regions, respectively. The red-bold line is from the $H_\mathrm{L}$ value with no contamination assumption, and two thin red lines are from its 2$\sigma$ upper and lower values.  The dashed and dotted lines are from the $H_\mathrm{L}$ values assuming that 20\% and 40\% of the excess flux is from contamination sources, respectively. The cyan hatched area represents the excluded region calculated from the 3$\sigma$ upper limit on $\pi_\mathrm{E}$.
}
\label{fig:ms_vs_DL}
\end{center}
\end{figure}

Using $\theta_\mathrm{E}$,  $m_\mathrm{host}$ is expressed by
\begin{eqnarray}
m_\mathrm{host} = \frac{1}{1+q}\frac{\theta_\mathrm{E}^2}{\kappa \pi_\mathrm{rel}},
\end{eqnarray}
where 
$\pi_\mathrm{rel} \equiv \mathrm{AU} (1/D_\mathrm{L}-1/D_\mathrm{S})$.
In Figure \ref{fig:ms_vs_DL}, we plot the $m_\mathrm{host}$--$D_\mathrm{L}$ relation for the $\theta_\mathrm{E}$ value derived in Section \ref{sec:theta*_w_spec}, fixing $D_\mathrm{S}$ at 9.1 kpc. 

The other $m_\mathrm{host}$--$D_\mathrm{L}$ relation can be derived from the lens flux. We first assume that the observed excess flux comes solely from the lens flux (i.e., there is no contamination from other sources).
Because the excess brightness is measured most precisely in the $H$ band among the three bands, we use $H_\mathrm{excess}$ to derive the physical parameters.
 The absolute $H$-band magnitude of the host star, $M_{H,\mathrm{L}}$,  is calculated using the $H$-band lens flux $H_\mathrm{L}$, here $H_\mathrm{L} = H_\mathrm{excess}$, as a function of $D_\mathrm{L}$ by the following equation,
\begin{eqnarray}
\label{eq:M_H}
M_{H,\mathrm{L}} = H_\mathrm{L} - A_{H,\mathrm{L}} - 5 \log \frac{D_\mathrm{L}}{10\mathrm{pc}},
\end{eqnarray}
where  $A_{H,\mathrm{L}}$ is the $H$-band extinction up to $D_\mathrm{L}$, for which we simply assume $A_{H,\mathrm{L}} = A_\mathrm{H} D_\mathrm{L}/D_\mathrm{S}$. For a given $D_\mathrm{L}$, $M_{H,\mathrm{L}}$ can be converted to $m_\mathrm{host}$ via an isochrone model. We use the PARSEC isochrones version 1.2S \citep{2012MNRAS.427..127B,2014MNRAS.444.2525C}, which was improved for low-mass stars over the previous versions. We assume the host star's age of 4 Gyr and a solar metallicity. 
In Figure \ref{fig:ms_vs_DL}, we include a plot of this $m_\mathrm{host}$--$D_\mathrm{L}$ relation. We also plot the 3$\sigma$ excluded area from the $\pi_\mathrm{E}$ upper limit calculated in Section \ref{sec:piE_limit} as the cyan hatched region.
As a result, we find that the two $m_\mathrm{host}$--$D_\mathrm{L}$ relations cross each other at $0.4 \lesssim D_\mathrm{L}$/kpc $\lesssim 2.6$ and $0.1 \lesssim m_\mathrm{host}$/M$_\odot \lesssim 0.55$ within 2 $\sigma$, indicating that the host star is likely a nearby M dwarf.

To properly estimate $m_\mathrm{host}$ and $D_\mathrm{L}$ as well as other related physical parameters including the planetary mass, $m_\mathrm{p}$, and the projected star--planet separation, $r_\perp$, we calculate probability distributions of these parameters by means of the Monte Carlo technique. Specifically, we solve for 
$m_\mathrm{host}$, $D_\mathrm{L}$, $m_\mathrm{p}$, and $r_\perp$ from a given set of the following observed and assumed parameters: {\boldmath$X$} = \{$\rho$, $q$, $s$, $\theta_*$, $H_\mathrm{L}$, $A_{H}$, $D_\mathrm{S}$, [M/H]\}, where [M/H] is the metallicity of the host star.
We repeat this calculation about 10$^5$ times by randomly sampling {\boldmath$X$}.
For the random distributions, we use the posterior distributions obtained from the MCMC analysis in Section \ref{sec:MCMC} for $\rho$, $q$, and $s$, and we use a Gaussian distribution for $\theta_*$, $H_\mathrm{L}$, $A_{H}$, $D_\mathrm{S}$, and [M/H]. We  assume [M/H]=-0.05 $\pm$ 0.2, which is consistent with the metallicity distribution of nearby M dwarfs \citep[e.g.][]{2014MNRAS.443.2561G}. Note that we fix the stellar age at 4~Gyr, because the age dependence on the isochrone models is negligible for moderately mature M dwarfs ($\gtrsim$0.5~Gyr).

\begin{figure*}
\begin{center}
\includegraphics[width=14cm]{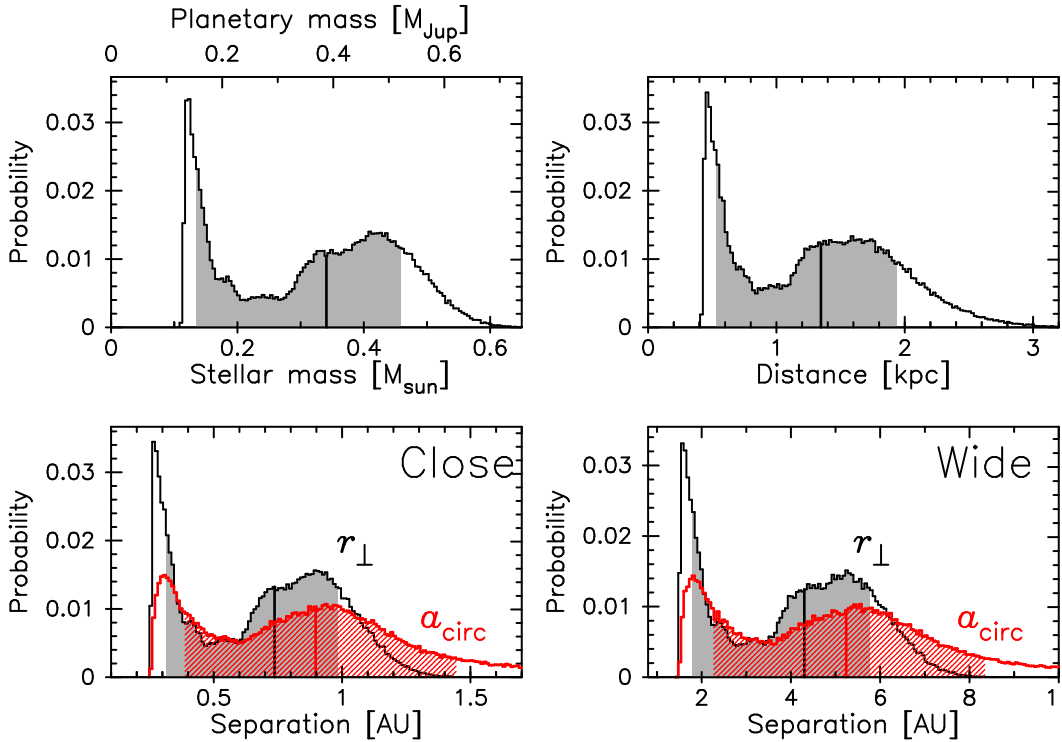}
\caption{
Probability distributions of the stellar mass (upper left), distance to the lens system (upper right), planet--star separation for the close model (lower left), and that for the wide mode (lower right). In the upper left  panel, the planetary mass converted by multiplying the stellar mass by $q=1.09 \times 10^{-3}$ is shown in the top horizontal axis. Probability distributions of the semi-major axis ($a$) assuming a circular orbit, random orbital inclination, and uniform occurrence frequency are indicated in red in the lower panels. In each panel, the vertical solid line indicates the median value and gray (red-shared for $a$) area represents the 1$\sigma$ confidence region.}
\label{fig:final_probability}
\end{center}
\end{figure*}

The probability distributions of the lens-system parameters are shown in Figure \ref{fig:final_probability}.
Each distribution shows a bimodal shape, which comes from the fact that the two $m_\mathrm{host}$-$D_\mathrm{L}$ relations form a Y-like shape. If $\theta_\mathrm{E}$ takes an upper-side value (above the black solid line), then the stellar mass will most likely be very low ($\sim1.2 M_\odot$), which forms the left-hand sharp peak. On the other hand, if $\theta_\mathrm{E}$ takes a lower-side value, then the stellar mass can take a value in a wider range of $\sim$0.3--0.5 $M_\odot$, which forms the right-hand loose peak. Therefore, this bimodal shape does not come from two distinct solutions but from one solution. For this reason, for a final value of each parameter and its uncertainties, we just take the median and 68.3 \% confidence intervals of each probability distribution.
As a result, we obtain $m_\mathrm{host}=$ \rev{0.34~$^{+0.12}_{-0.20}$}~M$_\odot$, $D_\mathrm{L} =$ 1.3~$^{+0.6}_{-0.8}$~kpc, $m_\mathrm{p}=$ \rev{0.39~$^{+0.14}_{-0.23}$}~M$_\mathrm{Jup}$ (\rev{123~$^{+44}_{-73}$}~M$_\mathrm{Earth}$), and $r_\perp=$ \rev{0.74~$^{+0.26}_{-0.42}$}~AU for the close model and \rev{4.3 $^{+1.5}_{-2.5}$}~AU for the wide model. In addition, we calculate the probability distribution of the de-projected semi-major axis $a$, assuming a circular orbit, uniform distribution of $a$, and random distributions of orbital inclination and phase \citep{1992ApJ...396..104G}, resulting in the estimated semi-major axis of $a_\mathrm{circ}=\rev{0.90^{+0.52}_{-0.50}}$ AU for the close model and $a_\mathrm{circ}=\rev{5.2^{+3.1}_{-2.9}}$ AU for the wide model.  All the resultant parameters are listed in Table \ref{tbl:physical_parameters}.

\begin{deluxetable*}{lcccc}
\tablecaption{\rev{Parameters of the Planetary System} 
\label{tbl:physical_parameters}}
\tablehead{
Parameters & Units & & Values &
}
\startdata
Contamination fraction $^{a}$ & \% & 0 ($<$10) & 20 (10--30) & 40 (30--50) \\[3pt]
\hline
Stellar mass ($m_\mathrm{host}$) & M$_\odot$ & 
  0.34 $^{+0.12}_{-0.20}$ &
  0.19 $^{+0.21}_{-0.07}$ &
  0.13 $^{+0.17}_{-0.02}$\\[3pt]
Planetary mass ($m_\mathrm{p}$) & M$_\mathrm{Jup}$ & 
  0.39 $^{+0.14}_{-0.23}$ &
  0.22 $^{+0.24}_{-0.08}$ &
  0.15 $^{+0.19}_{-0.02}$ \\[3pt]
& M$_\mathrm{\oplus}$ & 
  123 $^{+44}_{-73}$ &
  70 $^{+75}_{-26}$ &
  47 $^{+60}_{-7}$\\[3pt]
Distance ($D_\mathrm{L}$) & kpc & 
  1.3 $^{+0.6}_{-0.8}$ &
  0.85 $^{+0.93}_{-0.34}$ &
  0.63 $^{+0.83}_{-0.12}$\\[3pt]
Projected separation ($r_\perp$) & AU &&&\\
\hspace{12pt} Close model & & 
  0.74 $^{+0.26}_{-0.42}$ &
  0.45 $^{+0.44}_{-0.17}$ &
  0.33 $^{+0.39}_{-0.06}$\\[3pt]
\hspace{12pt} Wide model &  & 
4.3 $^{+1.5}_{-2.5}$ &
2.6 $^{+2.6}_{-1.0}$ &
1.9 $^{+2.3}_{-0.3}$\\[3pt]
Semi-major axis ($a_\mathrm{circ}$) & AU &&&\\
\hspace{12pt} Close model & & 
  0.90 $^{+0.52}_{-0.50}$ &
   0.67 $^{+0.55}_{-0.34}$&
  0.45 $^{+0.53}_{-0.14}$\\[3pt]
\hspace{12pt} Wide model & & 
5.2 $^{+3.1}_{-2.9}$&
3.9 $^{+3.2}_{-2.0}$&
2.6 $^{+3.1}_{-0.8}$\\[3pt]
\hline
Contamination probability & \% &&&\\
\hspace{12pt} Chance alignment star & & 
92.9 &
5.2 &
1.9\\[3pt]
\hspace{12pt} Companion to the source & & 
89.7 & 
6.7 & 
3.5 \\[3pt]
\hspace{12pt} Companion to the lens & & 
99.62 & 
0.22 & 
0.16 \\[3pt] 
\hspace{12pt} Total & & 
83.2 & 
11.4 &
5.4
\enddata
\tablenotetext{a}{\ The number in parenthesis is for the probability listed in the bottom part of this table.}
\end{deluxetable*}

\subsection{Contamination Scenarios}
\label{sec:contami}

If the measured excess flux is contaminated by extra sources, the actual lens flux would decrease by the contaminated flux, pushing down the lens mass. The amount of this decrement, however, is limited due to the existence of the upper limit of $\pi_\mathrm{E}$. In other words, there is an upper limit of the contamination flux; we estimate $\sim$50\% of the excess flux as this limit above which the two mass--distance relations, from $H_\mathrm{L}$ and from $\theta_\mathrm{E}$, do not cross each other outside the $\pi_\mathrm{E}$ excluding region within the uncertainties.
 We look into the effects and probabilities of contaminations in the following sections.

\subsubsection{Effects of Contamination} 
\label{sec:effects_of_contami}
First, we estimate how the contamination changes the physical parameters of the lens system. We assume two contamination levels, $f=$ 0.2 and 0.4, where $f$ is the ratio of the contamination flux to the excess flux, and calculate the posterior probabilities of the physical parameters for each contamination level in the same way as the previous section. The results are summarized in Table \ref{tbl:physical_parameters}. As the contamination level increases, the stellar (and planetary) mass and distance shrink to $\sim$1/3 and $\sim$1/2, respectively. Nevertheless, in many cases the 1$\sigma$ allowed regions of each parameter overlap each other; the result obtained in Section \ref{sec:no_contami}, that the lens system is a nearby M-dwarf planetary system, does not change.

\subsubsection{Chance-alignment Star}
\label{sec:alignment_star}
Next, we calculate the probabilities of contamination for each extra source and for each contamination level.
The first case of contamination source is an unrelated chance-alignment star.
On the $18'' \times 18''$ FOV of the IRCS/$H$-band image,
we detect 58 and 53 stars in the flux ranges for the contamination levels of $f=0.2$ (10\%--30\% of the excess flux) and $f=0.4$ (30\%--50\% of the excess flux), respectively. To account for the detection incompleteness, we embed a hundred of artificial stellar objects on the IRCS image with a flux corresponding to each contamination level, and try to detect them. We recover 29\% and 72\% of the embedded stars for the respective contamination levels, implying that there are potentially 197 and 72 stars on the IRCS image, respectively. Adopting a circle with an 8 pixel (0\farcs16) radius as the contamination cross section, we calculate that the chance-alignment probability for the contamination levels of $f=0.2$ and 0.4 are 5.2\% and 1.9\%, respectively.
The calculated probabilities are summarlized in Table \ref{tbl:physical_parameters}.

\subsubsection{Companion to the Source Star}
The second contamination source is a companion to the source star. 
Using Equation (\ref{eq:M_H}) and the PARSEC isochrones version 1.2S, we calculate that 10\%--30\%  and 30\%--50\% of the excess flux correspond to the companion mass of 0.60--0.83 M$_\odot$ and 0.83--0.96 M$_\odot$, respectively.
Using the binary (including triplet and more) fraction of the FGK stars of 46\% measured by \citet{2010ApJS..190....1R} and the companion-mass distribution around the FGK stars from the same paper, we estimate that the probabilities of a G-dwarf source having a companion with the respective mass ranges are 11\% and 5.8\%.
We further constrain these probabilities from the limits on the projected separation between the source and companion stars.  We set an upper limit of 0\farcs16 in the same way as the case of the chance-alignment star, while setting a lower limit of 1/4$\theta_\mathrm{E}=0.36$ mas \citep{2014ApJ...780...54B}, below which we would see an additional bump in the microlensing light curve. This angular range is converted to the semi-major axis range at 9.1 kpc of 4.0 -- 1780 AU, or the orbital-period range of $< \log P$ (days) $<$ 7.4. Adopting a log-normal distribution with the mean of $\log P$=5.03 and standard deviation of $\sigma_{\log P}$=2.28 as the orbital-period distribution of FGK-dwarf binaries \citep{2010ApJS..190....1R}, we estimate that the fraction of binaries within this orbital-period range is 61\%. Therefore, the total probabilities for the contamination levels of $f=0.2$ and 0.4 are 6.7\% and 3.5\%, respectively.

\subsubsection{Companion to the Lens Star}
The last case of the contamination source is a companion to the M-dwarf lens star.
In this case, the distance to the lens system and the masses of the hypothetical-binary components change depending on the contamination level $f$.
We therefore use the following equation to calculate the probability that the lens star has a companion:
\begin{eqnarray}
\label{eq:P_f}
P (f) = F_\mathrm{binary} \times  \sum_f F_{q_\mathrm{c}} (f) \times F_{a_\mathrm{c}} (f),
\end{eqnarray}
where $F_\mathrm{binary}$ is the fraction of M dwarfs that have a companion, and $F_{q_\mathrm{c}}$ and $F_{a_\mathrm{c}}$ are the fractions of M-dwarf companions that have the mass ratio $q_\mathrm{c}$ and  semi-major axis $a_\mathrm{c}$, respectively.

To calculate $F_{q_\mathrm{c}}$, we estimate the companion mass using Equation (\ref{eq:M_H}) and an isochrone model of the PHOENIX/AMES-dusty model \citep{2001ApJ...556..357A} for a given $f$. The companion mass varies in the range of 0.10--0.11 M$_\odot$ for $0.1 < f < 0.5$. Then, the host star's mass for a given $f$ is calculated in the same way as in Section \ref{sec:effects_of_contami} to derive $q_\mathrm{c}$. The $q_\mathrm{c}$ value varies in the range of 0.35--0.94 depending on $f$. For the $q_\mathrm{c}$ distribution, we simply assume a uniform distribution \citep{2014ApJ...789..102J}.
We note that because the mass distribution of the M-dwarf companion is not yet clear, the choice of this simple distribution might be systematics in the probability estimation. However, as shown below, the probability of contamination from a lens companion is much smaller than the other contamination sources, and therefore the choice of mass distribution should not affect the total contamination probability.

To calculate $F_{a_\mathrm{c}}$, we set the limits on the binary separation.
We set the same upper limit of 0\farcs16 as the previous cases, which is converted to $a_\mathrm{c}$=\rev{229} AU for $f$=0.1 and $a_\mathrm{c}$=\rev{120} AU for $f$=0.5.  For a lower limit, we consider a caustic induced by a hypothetical companion by which additional anomalies could be produced in the microlensing light curve. We adopt an upper limit on the full width of the hypothetical caustic as $w = 4q_\mathrm{c}/s_\mathrm{c}^2 < u_0 \sim 1.4 \times 10^{-3}$, where $s_\mathrm{c}$ is the projected separation of the companion and the host star normalized by $\theta_\mathrm{E}$. This inequality provides the lower limits of the angular separation of 46 mas for $f=0.1$ and 75 mas for $f=0.5$, which correspond to  $a_\mathrm{c}$=53--65 AU depending on $f$.  For the distribution of $a_\mathrm{c}$, we adopt a log-normal distribution with the mean of $\log a_\mathrm{c}$ (AU)= 0.80 and a standard deviation of $\sigma_{\log a_\mathrm{c}}$=0.48 \citep{2014ApJ...789..102J}. By summing up Equation (\ref{eq:P_f}) with a step size of $\Delta f=5$\%, we calculate the probabilities that the lens star has a companion as 0.22\% and 0.16\% for $f$=10\%--30\% and 30\%--50\%, respectively.

\subsubsection{Total Probabilities of Contamination}
We calculate that the total contamination probabilities considering the above three scenarios are 11.4\% and 5.4\% for $f=$10\%--30\% and 30\%--50\%, respectively. This means that the probability that the flux contamination fraction is less than 10\% and 30\% is 83.2\% and 94.6\%, respectively. 
Therefore, it is most likely that most of the excess flux comes from the lens star itself. We note that in the case of $f<10$\%, the 1$\sigma$ allowed ranges of the physical parameters of the lens system are entirely included in those for the non-contamination case. Hereafter we simply assume that there is no contamination, and take the parameter values calculated in Section \ref{sec:no_contami} as our final values.
Note that the possibilities of the first two contamination sources, the chance-alignment star and companion to the source star, can be tested in the future by spatially resolving the source and lens stars.

\begin{center}
\begin{figure}
\includegraphics[width=8cm]{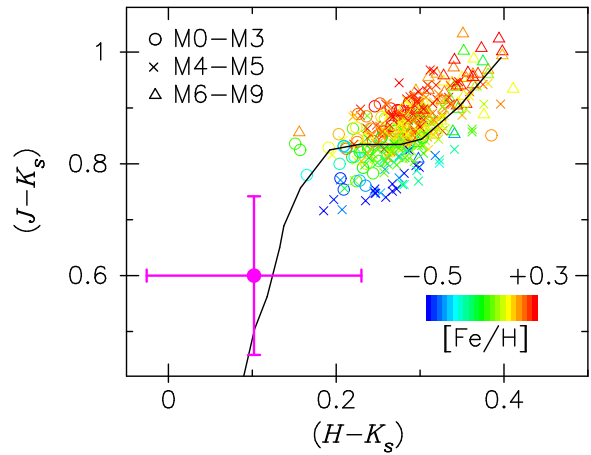}
\caption{
($H-K_\mathrm{s}$) vs ($J-K_\mathrm{s}$) for M dwarfs. The position of OGLE-2012-BLG-0563L is indicated by the magenta point with error bars. The open circles, triangles, and crosses are metallicity-measured nearby M dwarfs presented in \citep{2014AJ....147...20N} within the spectral sub-types of M0-M3, M4-M5, and M6-M9, respectively. Colors represent the metallicity ([Fe/H]). The black solid line denotes a main-sequence track of \citet{1988PASP..100.1134B}.
}
\label{fig:hkjk}
\end{figure}
\end{center}

\subsection{($J-K_\mathrm{s}$) and ($H-K_\mathrm{s}$) colors of the host star}
\label{sec:jkhk}

Because the $J$- and $K_\mathrm{s}$-band excess fluxes are also measured, the ($J-K_\mathrm{s}$) and ($H-K_\mathrm{s}$) colors of the host star can be estimated, which can be used as an independent check of whether the host star's mass derived in Section \ref{sec:no_contami} is consistent with an M dwarf.
We estimate the reddening for the host star by interpolating the online tables of \citet{2014A&A...566A.120S} for the Galactic coordinate and distance of the lens system, yielding $E(J-K_\mathrm{s})_\mathrm{L}=0.045\pm0.024$ and $E(J-H)_\mathrm{L}=0.011\pm0.007$, where  the allowed range of the distance is taken into account.
Then, assuming that the excess fluxes come solely from the host star,  we calculate the de-reddening colors as ($J-K_\mathrm{s}$)$_\mathrm{L}$=0.60 $\pm$ 0.14 and ($H-K_\mathrm{s}$)$_\mathrm{L}$=0.10 $\pm$ 0.13. We plot it on the color-color diagram in Figure \ref{fig:hkjk}, along with the distribution of the metallicity-measured nearby M dwarfs presented in \citet{2014AJ....147...20N} and a main-sequence track of \citet{1988PASP..100.1134B}. 
The position of the host star is not exactly at the majority of M dwarfs, but is more consistent with the region for K dwarfs. However, considering the large error bars, the measured colors are still marginally consistent with metal-poor mid-to-early M dwarfs.

We note that the spectral-type estimation from these color measurements is more sensitive to systematics than the mass measurement from $\theta_\mathrm{E}$ and $H_\mathrm{L}$. Although a systematical change of as large as 0.4 mag in $H_\mathrm{L}$ would keep the stellar mass within the M-dwarf range, only a 0.2 mag systematical shift in colors would easily change the inferred spectral type. In addition, potential systematics in the color measurements are larger than that in $H_\mathrm{L}$. Because there is no $J$- and $K_\mathrm{s}$-band light curves, $J_\mathrm{S}$ and $K_\mathrm{s, S}$ are estimated via several calibration processes, including the estimation of $A_J$ and $A_{K_\mathrm{s}}$, which could be a source of systematics in the colors (see Section \ref{sec:from_VH}). On the other hand, $H_\mathrm{S}$ is directly measured from the light curve, which minimizes the uncertainty in $H_\mathrm{L}$.
Therefore, it is more likely that the slight discrepancy in the inferred spectral type of the host star comes from systematics, or simply statistical errors, in the measured colors rather than those in the measured mass.
We further note that it could also be explained by a contamination of a distant early-type dwarf or a disk WD, which would cause a bluer shift in the measured colors. However, the probabilities of these scenarios should be much lower than those calculated in Section \ref{sec:alignment_star}, because the majority of possible chance-alignment stars are bulge dwarfs. Instead of identifying the cause of this discrepancy, we will discuss the prospects of improving the color estimation in Section \ref{sec:discussion}.

\begin{center}
\begin{figure}
\includegraphics[width=9cm]{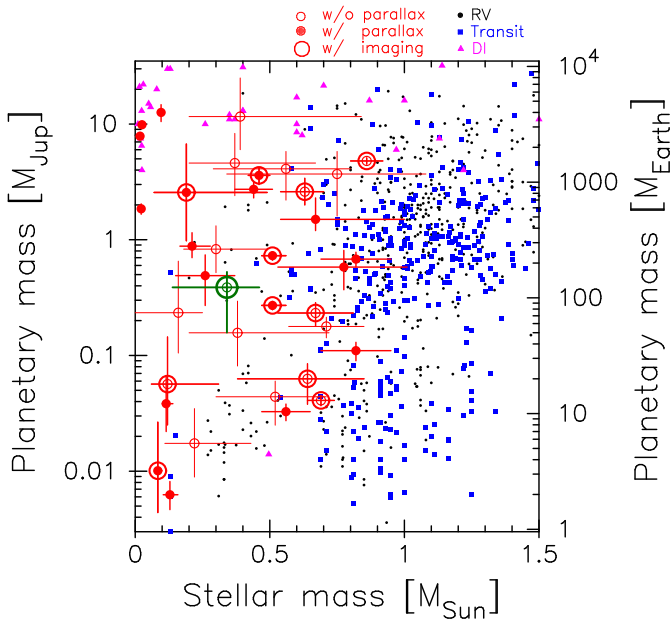}
\caption{Distribution of known exoplanets in the planetary-mass vs. stellar-mass plane. The black dots, blue squares, magenta triangles, and red circles are planets observed by the radial velocity, transit, direct imaging, and microlensing methods, respectively. The values of microlensing planets are  from literature, while those of the others are from http://exoplanet.eu. Microlensing planets with and without parallax measurements are denoted as open and filled circles, respectively, and those for which high-resolution imaging was used to constrain the masses are indicated by large open circles. The OGLE-2012-BLG-0563L system is indicated by green. 
\label{fig:msmp}
}
\end{figure}
\end{center}
\begin{figure}
\begin{center}
\includegraphics[width=8cm]{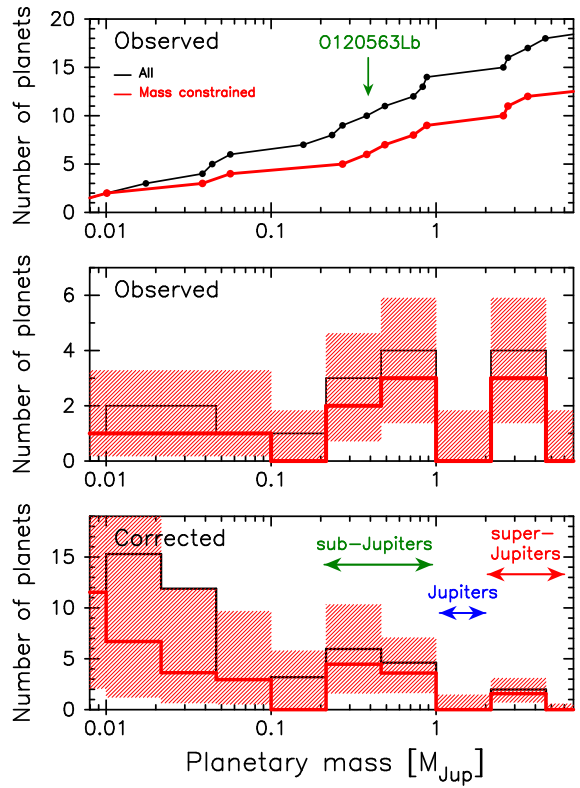}
\caption{\rev{Mass distribution of microlensing planets hosted by M dwarfs ($0.08 < m_\mathrm{host}/M_\odot < 0.55$). (Top) a cumulative distribution. The thin black line and bold red line are for all planets and for the planets with the mass constrained by parallax and/or high-resolution imaging, respectively. The location of OGLE-2012-BLG-0563Lb is indicated by an arrow. (Middle) the same as the top panel, but a histogram. The shaded area indicates the 68\% poisson confidence region for the histogram of the mass-constrained planets. \textcolor{red}{Note that the lower limits shown in the published version are erroneous, which are corrected in this figure.}  (Bottom) the same as the middle panel, but the number of planets per bin is corrected by a scaled detection efficiency of $(q/q_\mathrm{Jup})^{0.6}$, where $q$ is the planet--star mass ratio and $q_\mathrm{Jup}=M_\mathrm{Jup}/0.35M_\odot$.} \label{fig:mp_hist}}
\end{center}
\end{figure}

\section{Discussion}
\label{sec:discussion}

The derived lens parameters indicate that the lens system consists of an M dwarf orbited by a Saturn-mass planet.
Although the planet--star separation is not uniquely constrained due to the wide-close degeneracy, both solutions locate the planet at cold regions with the estimated semi-major axis close to (for the close model) or well beyond (for the wide model) the snow line, which we estimate to be $\sim$1 AU. Therefore, the planet is located in the region where the most microlensing planets have been discovered.

In Figure \ref{fig:msmp}, we show the distribution of known exoplanets in the stellar-mass versus planetary-mass plane. 
OGLE-2012-BLG-0563Lb is the fifth sub-Jupiter-mass ($0.2 \lesssim m_\mathrm{p}/M_\mathrm{Jup} \lesssim 1$) microlensing planet around M dwarfs ($0.08 \lesssim m_\mathrm{host} \lesssim 0.55 $M$_\odot$) with the mass constrained by either parallax or high-resolution imaging, following OGLE-2006-BLG-109Lb, c \citep{2008Sci...319..927G,2010ApJ...713..837B}, OGLE-2011-BLG-0251Lb \citep{2013A&A...552A..70K}, and OGLE-2011-BLG-0265Lb \citep{2014arXiv1410.8252S}. With this new discovery, 
it has become clear that these sub-Jupiter class planets form a population
around M dwarfs. In contrast, $\sim$1--2 Jupiter-mass planets are relatively rare around the same type of star.
This trend is strengthened when considering the fact that the detection efficiency is higher for more massive planets. 

To make this trend more clear, we show the mass distribution of known microlensing planets around M dwarfs in Figure \ref{fig:mp_hist}. The upper panel shows a cumulative distribution function of the observed planets, with a black line for all planets and red one for those with a mass constrained by parallax and/or high-resolution imaging. The middle panel shows a histogram for the same sample, where the red shaded region indicates 1$\sigma$ (68\%) Poisson uncertainty for the mass-constrained sample. The lower panel shows a ``corrected'' histogram for which the number of planets per one detection is corrected by an approximate relative detection efficiency  of $(q/q_\mathrm{Jup})^{\alpha}$, where $q$ is the planet--star mass ratio and $q_\mathrm{Jup}=M_\mathrm{Jup}/0.35M_\odot$. We adopt $\alpha=0.6$ for all planets as a mean value for the central-caustic and planetary-caustic events \citep{2010ApJ...710.1641S}. In this histogram, the possible paucity of Jupiters compared to sub-Jupiters is marginally seen at the 2 $\sigma$ level. 
We note that similar trends have also been observed in RV surveys \citep[e.g.,][]{2013A&A...549A.109B} and the {\it Kepler} survey \citep[e.g.,][]{2012ApJS..201...15H}, although the orbital regions probed by these surveys are much inner than those probed by microlensing.
If this trend is true, then it could be a consequence of the planetary formation process in the core-accretion scheme, in which gas-accreting planets around M dwarfs can rarely reach a Jovian mass  due to the lack of solid and gas materials \citep[e.g.,][]{2005ApJ...626.1045I,2011A&A...526A..63A}. On the other hand, several super-Jupiter-mass planets ($m_\mathrm{p} \gtrsim 2 M_\mathrm{Jup}$) have also been discovered around M dwarfs by both microlensing and direct imaging. These planets have challenged the core-accretion scenarios, and might have formed by other mechanisms, such as the gravitational instability \citep[e.g.,][]{2006ApJ...643..501B}. 
We note that the steepness of planetary-mass function around low-mass stars has already been pointed out by several microlensing studies \citep{2010ApJ...720.1073G,2010ApJ...710.1641S,2012Natur.481..167C,2014ApJ...791...91C}, however, these studies applied a simple mass function with a constant slope over all mass range \citep{2010ApJ...720.1073G,2010ApJ...710.1641S,2012Natur.481..167C}, or coarsely classified gas planets, treating those with a mass ranging  $1 < m_\mathrm{p}/M_\mathrm{Jup} < 10$ as one group \citep{2014ApJ...791...91C}. We  instead demonstrate that it has become possible to discuss a finer structure of mass function with increasing the number of planet discoveries.

To further clarify the planetary formation process around M dwarfs, further increasing the statistics of microlensing planets in terms of their number and accuracy is required. To this end, continuous efforts of not only photometric surveys/follow-ups but also high-resolution imaging are important. 
Indeed, OGLE-2012-BLG-0563 is the first M-dwarf-host planetary event without parallax for which the lens flux is clearly detected, demonstrating that ground-based near infrared (NIR) AO imaging can play a crucial, complementary role to constrain the mass of the M-dwarf-host planetary systems.

Another important aspect of AO imaging in the NIR is that it can in principle provide information about the metallicity of microlensing M-dwarf host stars from the ($J-K_\mathrm{s}$) and ($H-K_\mathrm{s}$) colors. 
Giant planet-hosting sun-like stars tend to be metal rich, which has been cited as strong evidence of the core-accretion scenarios \citep[e.g.,][]{2004A&A...415.1153S,2005ApJ...622.1102F}. Probing the metallicity of microlensing M-dwarf host stars orbited by giant planets can thus be an important tool to test the planetary formation models of cold gas giants around M dwarfs. As shown in Section \ref{sec:jkhk}, the measured ($J-K_\mathrm{s}$) and ($H-K_\mathrm{s}$) colors of OGLE-2012-BLG-0563L are marginally consistent with a low metallicity mid-to-early M dwarf;  if this is true, then this would imply that the discovered planet could be a rare sample being formed by the core-accretion model, or could have formed via other formation mechanisms. However, the current error bars for these colors are too large to be conclusive.
These uncertainties mainly come from the several calibration steps to derive the source magnitudes, in particular $J_\mathrm{S}$ and $K_\mathrm{s,S}$, which are not derived from light curves but are estimated via the temperature-metallicity-color conversions and the extinction laws.
As described in Section \ref{sec:jkhk}, these calibration processes could also be a source of systematics.

These uncertainties can be reduced by future observations. Because our IRCS/$JHK'$-band observation was conducted at the time when the source star was still magnified by a factor of 1.47, a second epoch observation (at the baseline) with the same instrument/filters will provide $JHK'$-band ``light curves''. These light curves can directly provide the instrumental ($J-K'$) and ($H-K'$) colors of the host star, which can then be converted to the colors in the 2MASS system with only color-color corrections. This additional AO observation will provide a better estimation of the host star's colors while avoiding the calibration steps for the source star. Ultimately,  one can obtain a further better estimation of the host star's color by spatially resolving the host star from the source star, thus removing the background ``noise''.
This observation will be possible in 10 years after the event peak with the current ground-based facilities.
Note that such a spatially resolving imaging will also be able to improve the $\theta_\mathrm{E}$ estimation, and hence to refine the physical parameters of the lens system, as was done in \citet{2015ApJ...808..170B} and \citet{2015ApJ...808..169B}. This observation will be possible in four years, from the event peak, if {\it the Hubble Space Telescope} is used.

\section{Summary}
\label{sec:summary}

We present the discovery of a microlensing planet OGLE-2012-BLG-0563Lb, which was detected through intensive photometric observations of a high-magnification event. A light curve analysis clearly detects the planetary signal of $q \sim 10^{-3}$  with $\Delta \chi^2 > 1000$. On the other hand, we do not detect a clear parallax signal in the light curve; we only place an upper limit on $\pi_\mathrm{E}$, preventing us from deriving absolute physical parameters of the lens system from the light curve alone.

Thanks to the spectral information of the source star obtained by  \citet{2013A&A...549A.147B} at a high-magnification state, we derive the source's intrinsic color and magnitude with a minimum assumption about the dust extinction and reddening, and obtain a better estimation of the source's angular radius $\theta_*$ as well as the Einstein radius $\theta_\mathrm{E}$. We also estimate the source star's distance to be $D_\mathrm{S} = 9.1^{+0.9}_{-1.1}$ kpc, from the same spectral information.

To alternatively constrain the physical lens parameters,
we conducted a high-resolution $JHK'$-band imaging by using the {\it Subaru}/AO188 and IRCS instruments at the time when the source star was still magnified by a factor of 1.47. We successfully detected the excess flux from the host star on the source star position, allowing us to derive the absolute physical parameters of the lens system by combining it with $\theta_\mathrm{E}$ estimation and the upper limit on $\pi_\mathrm{E}$.
We find that the lens system is located at 1.3 $^{+0.6}_{-0.8}$ kpc from us, and consists of an M dwarf (\rev{0.34 $^{+0.12}_{-0.20}$}~M$_\odot$) orbited by a Saturn-mass planet (\rev{0.39 $^{+0.14}_{-0.23}$}~M$_\mathrm{Jup}$) at the projected separation of 0.74 $^{+0.26}_{-0.42}$ AU (close model) or 4.3 $^{+1.5}_{-2.5}$ AU (wide model). 
The probability of contamination in the measured host star's flux, which would reduce the stellar and planetary masses by a factor of up to three, is estimated to be  17\%. This possibility can be tested by future high-resolution imaging.

This is the fifth sub-Jupiter-mass microlensing planet around an M dwarf with the mass constrained by parallax and/or imaging. The relatively rich harvest of sub-Jupiters around M dwarfs is contrasted with a possible paucity of $\sim$1--2 Jupiter-mass planets around the same type of star.
This trend could be a consequence of the planetary formation process in the core-accretion scheme.  

This is also the first M-dwarf-host planetary event without parallax for which the lens flux is clearly detected, demonstrating that ground-based AO imaging can play a crucial role to constrain the mass of the M-dwarf-host planetary systems.
In addition, we show that NIR AO imaging can, in principle, constrain the metallicity of a microlensing M-dwarf host star from ($J-K_\mathrm{s}$) and ($H-K_\mathrm{s}$) colors. Although the current data only marginally prefer a low metallicity of OGLE-2012-BLG-0563L, further observations will be able to meaningfully constrain the metallicity and provide a new insight into the formation scenario of this planet.

\acknowledgements
This paper is based in part on data collected at {\it Subaru} Telescope, which is operated by the National Astronomical Observatory of Japan.
A.F. thanks Oscar A. Gonzalez for kindly providing VVV data.
A.F. was supported by the Astrobiology Project of the Center for Novel Science Initiatives (CNSI), National Institutes of Natural Sciences (NINS) (Grant Number AB261005).
T.S. acknowledges the financial support from the JSPS, JSPS23103002,JSPS24253004, and JSPS26247023. 
%% MOA
The MOA project is supported by grants JSPS25103508 and 23340064.
NJR is a Royal Society of New Zealand Rutherford Discovery Fellow.
%% uFUN
Work by C.H. was supported by Creative Research Initiative 
Program (2009-0081561) of National Research Foundation of Korea.
\proof{SD is supported by ``the Strategic Priority Research Program- The
Emergence of Cosmological Structures'' of the Chinese Academy of
Sciences (grant No. XDB09000000).}
%% OGLE
\rev{The OGLE project has received funding from the National Science Centre,
Poland, grant MAESTRO 2014/14/A/ST9/00121 to AU.}
%% RoboNet
C.S. received funding from the European Union Seventh Framework Programme (FP7/2007-2013) under grant agreement no. 268421.
K.A., D.M.B., M.D., K.H., M.H., C.L., C.S., R.A.S. and Y.T. would like to thank the Qatar Foundation for support from QNRF grant NPRP-09-476-1-078. 

Facilities: \facility{{\it Subaru}(IRCS)}, \facility{FTN}, \facility{FTS}, \facility{Liverpool:2m}, \facility{LCOGT}.

\bibliographystyle{apj}
\bibliography{ref}

\if0

\fi

%\bibliographystyle{apj}
%\bibliography{ref}

\end{document}